# Nature of the interactions determining the stacking motif of covalent organic frameworks

*Christian Winkler[1] and Egbert Zojer[1,]\**

[1] Institute of Solid State Physics, NAWI Graz, Graz University of Technology, Petersgasse 16, 8010 Graz, Austria

Covalent organic frameworks (COFs) have attracted significant attention due to their chemical versatility combined with a significant number of potential applications. Of particular interest are two-dimensional COFs, where the organic building units are linked by covalent bonds within a plane. Most properties of these COFs are determined by the relative arrangement of neighboring layers. These are typically found to be laterally displaced, which, for example, reduces the electronic coupling between the layers. In the present contribution we use dispersion-corrected density-functional theory to elucidate the origin of that displacement, showing that the common notion that the displacement is a consequence of electrostatic repulsions of polar building blocks can be misleading. For the representative case of COF-1 we find that electrostatic and van der Waals interactions would, actually, favor a cofacial arrangement of the layers and that Pauli repulsion is the crucial factor causing the serrated AA stacking. A more in-depth analysis of the electrostatic contribution reveals that the "classical" Coulomb repulsion between the boroxine building blocks of COF-1 suggested by chemical intuition does exist, but is overcompensated by attractive effects due to charge-penetration in the phenylene units. The situation becomes more involved, when additionally allowing the interlayer distance to relax for each displacement, as then the different distance-dependences of the various types of interactions come into play. The overall behavior calculated for COF-1 is recovered for several additional COFs with differently sized π-systems and topologies, implying that the presented results are of more general relevance.

# 1. INTRODUCTION

Covalent organic frameworks (COFs) are highly porous crystalline materials consisting of organic building blocks linked by covalent bonds.[1–7] Because of their tunable structures, COFs have significant potential for various applications like gas storage,[8–10] gas separation,[11–14] catalysis,[15–17] energy storage,[18–20] or optoelectronics.[21–27] Among the different topologies of COFs, two-dimensional (2D) systems have received particular attention. Here, the individual organic building units are linked via covalent bonds within a plane, forming highly regular 2D layers. These layers then stack on top of each other and the resulting stacks are held together primarily by comparably weak van der Waals interactions. Important for the properties (electronic, optical, and catalytic) of the resulting three-dimensional (3D) stacks is the packing motif of consecutive 2D layers,[7,26,28–33] as it defines the shape of the pores and the overlap of the π-systems of neighboring sheets. The latter is crucial for the electronic structure of the system, as depending on the symmetry and nodal structure of the involved orbitals/wavefunctions, the resulting systems can be insulating, semiconducting, or even metallic.[34,35]

The vast majority of the reported 2D COFs exhibit either eclipsed (cofacial) or serrated (shifted) AA stacking,[28,30,36–41] where the actual magnitude of the shift is hard to determine experimentally via x-ray diffraction due to the large peak broadening in the typically investigated powder samples.[26,36,42] Different stacking motifs are found, for example, when the 2D sheets are not entirely planar.[38] In the following we will , however, focus on planar COFs, as these systems allow a more straightforward analysis of the interplay between different geometric degrees of freedom and the energetic stability of the respective COF. Of particular appeal for such an analysis are COF-1 and COF-5 (the structures first reported in ref 2). As far as modelling studies on these systems are concerned, Zhou et al.[36] explicitly showed (employing density functional tight-binding methods) that the total energy of the COF becomes a minimum for shifted layer arrangements with displacements of around 1 Å. They hypothesized that the alignment of neighboring π-orbitals plays an essential role for these shifts and compared the stacking motif of the aromatic rings to the situation in graphite, albeit without determining the nature of the interactions enforcing the serrated structure.[36] Lukose et al.[37] also performed a computational study on the alignment of layered COFs, again considering COF-1 and COF-5. These authors also identified similar shifts of consecutive layers (~1.4 Å) as the energetically favorable layer arrangements. In this work, as well as in Ref. 43, the authors argued that repulsive Coulomb interactions between B and O linking units in neighboring layers would cause the eclipsed AA stacking to be energetically unfavorable, but again without quantifying these interactions. Thus, to the best of our knowledge, a quantitative assessment of the different types of interactions as a function of the alignment of consecutive layers is still lacking. This lack prevents a fundamental understanding of the factors determining the packing motif in 2D COFs and also hinders the development of strategies for tuning the stacking arrangements of COFs and their resulting electrical, optical, and catalytic properties.[28,29,31] To generate such an understanding, in the present study we employ dispersion corrected density functional theory (DFT) calculations with a focus on decomposing the interlayer interactions in the prototypical model system COF-1 (see Figure 1 and ref 2 for the structure of this COF) into physically well-defined contributions arising from dispersion forces, electrostatic interactions, and exchange repulsions with orbital rehybridizations. To demonstrate the wider applicability of our findings, we eventually extend our analysis to COF-5 and COFs comprising porphyrin (Por-COF)[44] and hexabenzocoronene cores (HBC-COF).[39] The details of the structures of these COFs will be discussed below.

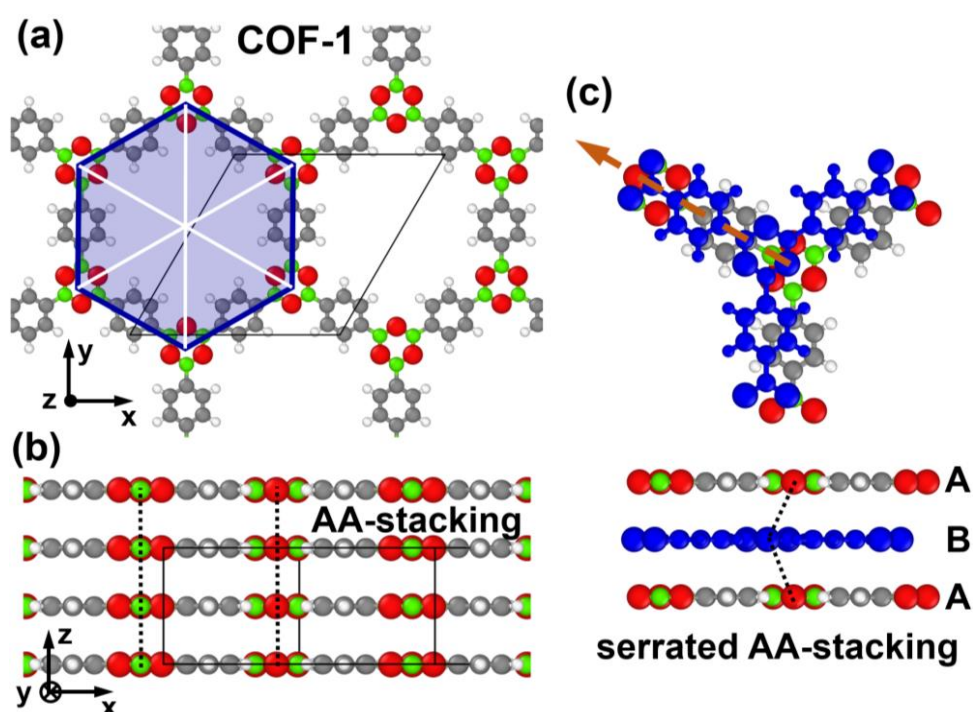

*Figure 1. Structure of COF-1 and the primarily considered shift direction: Panels (a) and (b) show top and side views of the structure of COF-1 for cofacial AA-stacking. The considered unit cell is shown by the thin, black, solid lines. The blue hexagon and the white lines highlight the hexagonal symmetry of the pores. In panel (c), a shifted (serrated) arrangement with a displacement of 1.75 Å parallel to one of the walls of the pore (direction **1**) is shown. The shift direction is shown as a dashed, orange arrow and the shifted layer is marked in blue. The*

*dashed black lines in panels (b) and (c) connect equivalent atoms in consecutive layers, indicating the stacking motif. Color code of the atoms: C ... grey, H ... white, B ... green, O ... red*

## 2. METHODS

For the investigations of the layered COFs considered in this study we employed dispersion corrected density functional theory, DFT, as implemented in the FHI-aims code.[45,46] For these calculations the PBE functional[47,48] was used and van der Waals interactions were considered by using the Tkatchenko-Scheffler,[49] TS, scheme. For comparison, also test calculations employing the computationally more costly many body dispersion correction,[50–52] MBD, were performed. The electronic band structure of COF-1 was also calculated using the HSE06[53,54] hybrid functional based on the PBE geometries. The corresponding data are shown in the Supporting Information. We used the conventional "tight" basis functions of FHI-aims with details for each atomic species described in the Supporting Information. For all bulk systems of the COFs a grid consisting of 3x3x6 k-points was employed for sampling reciprocal space, unless stated otherwise. Tests show that the total energy for this grid is well converged to within less than 1 meV. To describe the occupation of the electronic states a gaussian type smearing function with a width of 0.01 eV was employed. For geometry relaxations, the positions of individual atoms were allowed to relax until the largest force component on any of the atoms was below 0.01 eV/ Å.

The calculations of the potential energy surface for COF-1 as well as for shifts parallel to one of the pore walls were performed on structures employing the experimental lattice constants reported in literature (a=b=15.420 Å and c=3.328 Å).[2] The unit cell of COF-1 was constructed such that it contains two layers in stacking direction (layers A and B, see Figure 1). This allows displacing these layers along directions parallel to the xy-plane. In addition to considering systems with constant unit-cell height (and , thus, constant interlayer stacking distances) we also studied systems for which the unit-cell height was optimized for each displacement. Further details on the geometry relaxation and the determination of the optimal unit-cell heights and interlayer stacking distances can be found in the Supporting Information.

For the additional COFs considered in this work (COF-5, Por-COF, and HBC-COF), the in-plane lattice parameters had to be optimized, as for some of them no literature values are available. To find the optimal stacking arrangement in terms of in-plane shifts and (shift-dependent) stacking distance, we performed full geometry relaxations for the resulting bulk systems. Details on these simulations are provided in the Supporting Information together with test calculations for COF-1 in which the in-plane lattice parameters were also optimized. The latter yields somewhat smaller lattice constants than in the experiments (in good agreement with previous computational studies[55]) but does not significantly impact the results.

To calculate the full potential energy surface for lateral displacements between layers in COF-1 we employed Gaussian process regression, GPR, as implemented in scikit-learn.[56] The model vector consisted of the x and y positions of the shifted layer. As Kernel functions we used a combination of a constant kernel with the radial basis function kernel (RBF). To obtain the ideal hyper-parameters, the marginal log likelihood was maximized. The model was initially trained with 80 randomly chosen data points (i.e., displacements). Then 39 additional points were included at the xy positions of the maximum model uncertainty. The final model uncertainty was estimated to be well below 10 meV for displacements smaller than 3.5 Å and to be below 50 meV for shifts around 6 Å. Details of the model and the obtained model error are reported in the Supporting Information.

To determine the individual contributions to the interaction energy of the COFs, the system was split into two fragments, which consist of only one of the two layers in the unit cell each (layers A and B in Figure 1). Technically, this was achieved by removing one of the layers from the unit cell, while not changing the unit cell dimensions and the positions of the atoms in the other layer. Then the total energies of the COF containing both layers in the unit cell and of COFs comprising either only layers of type A or of type B were calculated. The resulting interaction energy between the fragments (i.e., layers) is then given by

$$\Delta E_{int} = E_{total}^{AB} - (E_{total}^{A} + E_{total}^{B}) \qquad (1).$$

This interaction energy can be decomposed into the interaction energy resulting from the PBE calculations and the contribution due to the a posteriori correction for (long range) van der Waals interactions

$$\Delta E_{int} = \Delta E_{PBE} + \Delta E_{vdW} \qquad (2).$$

The individual contributions can be readily obtained from the FHI-aims output for the full system and the two sub-systems as

$$\Delta E_{PBE} = E_{PBE}^{AB} - (E_{PBE}^{A} + E_{PBE}^{B}) \qquad (3a)$$

and

$$\Delta E_{vdW} = E_{vdW}^{AB} - (E_{vdW}^{A} + E_{vdW}^{B}) \qquad (3b)$$

A more involved step is to decompose the PBE interaction energy, $\Delta E_{PBE}$, into the electrostatic contribution due to the Coulombic interactions between the nuclei and electron clouds of the subsystems and in contributions due to exchange interactions and orbital rehybridizations. Various decomposition schemes that serve this purpose are available for finite-size systems, but for extended solids described by periodic boundary conditions such approaches are, unfortunately, rare. Therefore, a custom decomposition scheme was implemented as a post-processing tool for FHI-aims. This scheme is largely based on the periodic energy decomposition analysis (pEDA) scheme developed by Raupach and Tonner.[57,58] Within this scheme, the authors basically extended the energy decomposition analysis (EDA) method developed by Ziegler/Rauk[59,60] and Morokuma[61] to periodic boundary conditions. The key idea in this method is to partition the interaction energy $\Delta E_{int,elec}$ into well defined terms as shown in equation 4.

$$\Delta E_{int,elec} = \Delta E_{elstat} + (\Delta E_{Pauli} + \Delta E_{orb}) \qquad (4)$$

As a first step one can evaluate $\Delta E_{elstat}$ by considering the charge densities of the individual, non-interacting fragments A and B and use them to construct a combined system {A,B}. This combined system contains the charge densities of the non-interacting fragments A and B at the positions these fragments exhibit in the combined system. Consequently, the sum of the non-distorted charge densities $n_A$ and $n_B$ is used to describe the combined system. The energy of system {A,B} can then be calculated by performing a single shot DFT calculation without a self-consistency cycle. This calculation yields the electrostatic energy $E_{elstat}^{\{A,B\}}$ of the system as constructed from the fragments. The difference between this energy and the electrostatic energies of the individual fragments then yields the quasiclassical electrostatic interaction between the layers, $\Delta E_{elstat}$, as:

$$\Delta E_{elstat} = E_{elstat}^{\{A,B\}} - E_{elstat}^{A} - E_{elstat}^{B}. \qquad (5)$$

With the knowledge of $\Delta E_{elstat}$ it is possible to assess, whether an electrostatic repulsion between the (unperturbed) charge densities of consecutive layers is actually responsible for the common appearance of shifted (serrated) structures of 2D COFs. Another consequence of the overlap between the charge densities of the interacting sub-systems is Pauli repulsion, which is strongly repulsive. Additionally, the wavefunction overlap triggers orbital rehybridization, which lowers the energy of the entire system. This effect is, however, comparably small in stacked π-systems between which no interlayer bonds are formed.[62] In fact, Pauli repulsion and orbital rehybridization are intimately related, with a sizable part of the stabilizing effect of orbital rehybridization arising from a reduction of Pauli repulsion, especially in the absence of covalent interactions. Thus, in the following both energy contributions will be combined into a single term, $\Delta E_{orb,Pauli}$, which can be calculated from the overall interaction energy via:

$$\Delta E_{int} = \Delta E_{elstat} + \Delta E_{orb,Pauli} + \Delta E_{vdW}$$

## 3. RESULTS AND DISCUSSION

For layered 2D COFs, the two parameters characterizing the relative arrangement of the layers are (i) the interlayer stacking distance and (ii) the direction and magnitude of the shift between consecutive layers parallel to the plane of these layers. Both factors play a significant role in determining the actual properties of a COF. An advantage of computer simulations is that they allow varying both parameters independently. In particular, one can first address the question, how shifts between consecutive layers impact the energetic stability and the properties of 2D COFs for a fixed interlayer distance. In a second step one can then address the question to what extent the situation is altered when the interlayer distance is allowed to adapt for each shift. In the following, for both scenarios the focus will be on analyzing the impact of the stacking geometry on the interaction energy split into contributions from van der Waals attraction, Coulomb interactions, and the impact of orbital rehybridization and Pauli repulsion. As far as the impact of the relative arrangement of neighboring layers on COF properties are concerned, we will restrict the analysis to the electronic structure of the COF manifested in its electronic band structure.

### 3.1 Constant Interlayer Stacking Distance

Before displacing consecutive layers in specific directions, it is important to identify the directions in which minima of the potential energy surface are to be expected. The corresponding potential energy surface, PES, for the stacking distance fixed to the experimental literature value (3.328 Å) is shown in in Figure 2. This PES has been obtained employing Gaussian process regression, GPR, as described in the Methods section and in the Supporting Information. It shows the expected six-fold symmetry. Interestingly, the cofacial arrangement of successive layers (Δx = Δy = 0.00 Å) is particularly unstable. The global minima of the PES are found for shift directions parallel to the pore walls (see Figure 1) at displacements of around 1.75 Å. Therefore, in the following analysis we will focus on shifts along this direction. For the sake of comparison, we also calculated the total energy for a shift perpendicular to the pore wall (see Supporting Information) with the obtained data fully supporting the outcomes of the GPR fit.

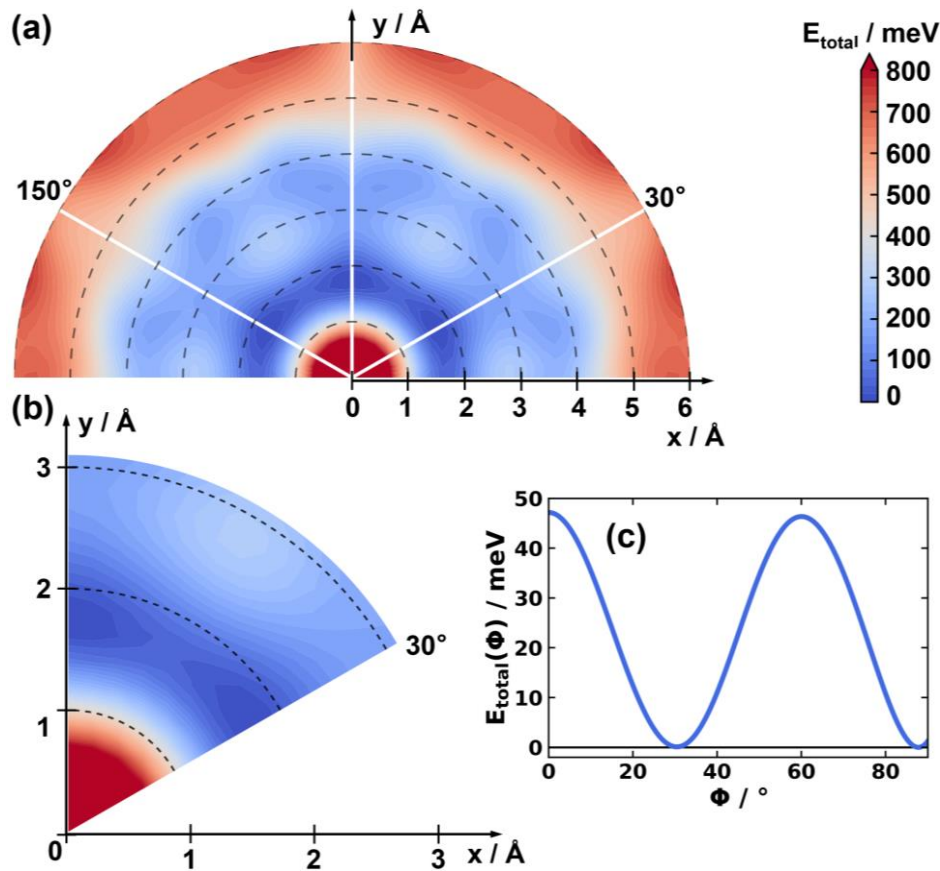

*Figure 2.* Panel (a) shows the total energy as a function of the displacements of consecutive COF-1 layers, i.e. the potential energy surface. The x- and y-axes are aligned in the same way as in Figure 1. The layers are displaced along directions parallel to the xy-plane. The obtained values of the total energy are specified per unit cell containing two COF layers and are reported relative to the energy of the global minimum structure. The hexagonal symmetry of the system is indicated by the white lines (see also Figure 1). These white lines also indicate directions parallel to the pore walls. Panel (b) shows a zoom into the region of smaller displacements for one of the symmetry inequivalent sections. Panel (c) shows the total energy as a function of the azimuthal angle angle $\phi$ (measured relative to the x-axis) and for a constant radius set to 1.75 Å, i.e. the value at which the minimum of the energy is observed in panels (a) and (b).

The evolution of the interaction energy, $\Delta E_{int}$, for the shift along one of the pore wall is shown in Figure 3a. It is plotted relative to the energy of the cofacially aligned layers, which amounts to -1290 meV per unit cell (see caption of Figure 3). In passing we note that the evolution of $\Delta E_{int}$ exactly follows that of the total energy, which is a consequence of its definition in Equ. (1) and the fact that the energies of the individual segments A and B for a fixed interlayer distance are independent of the shift. Notably, the value of $\Delta E_{int}$ is highest (least negative) for the cofacial arrangement, which confirms the notion from the GPR data that this is a particularly unstable structure. Likewise, at a displacement of around 1.75 Å the interaction energy displays a pronounced minimum and rises again for larger displacements. This behavior is consistent with the observations reported in literature,[36,37] although in our investigations the minimum occurs at slightly larger displacements. To understand the origin of that trend, the interaction energy is decomposed into contributions from van der Waals interactions, $\Delta E_{vdW}$, electrostatic interactions, $\Delta E_{elstat}$, and interactions due to Pauli repulsion and orbital rehybridization, $\Delta E_{Pauli,orb}$, where all energies in Figure 3a are plotted relative to their values for zero displacements (listed in the figure caption).

$\Delta E_{vdW}$ is most attractive for the cofacial arrangement (-3547 meV per unit cell), and then increases (i.e., becomes less negative) upon displacing the layers. This behavior is not unexpected, as by displacing the layers parallel to a pore wall, parts of the layers are moved over open pore space. This increases the average interatomic distances and causes a drop in vdW attraction. This effect becomes particularly pronounced for displacements above ~1.5 Å. Important to add is that we do not observe a dependence of this effect on the used van der Waals correction scheme, as is shown in the Supporting Information for MBD energy corrections instead of the otherwise employed TS scheme (see methods section).

A similar trend compared to $\Delta E_{vdW}$ is observed for the electrostatic energy shown by the blue open triangles in Figure 3a, where it has to be stressed that also $\Delta E_{elstat}$ remains negative (i.e., attractive) for all displacements with $\Delta E_{elstat}$ amounting to -1344 meV for the cofacial case. As far as the evolution of $\Delta E_{elstat}$ is concerned, it remains essentially constant for displacements up to around 2 Å and then becomes less negative (i.e., less attractive) for larger displacements. This means that the electrostatic attraction would be strongest in the cofacial case, and, thus, cannot be responsible for the serrated structure of COF-1, as hypothesized in refs 37,43. This raises the question as to the origin of the overall attractive nature of electrostatic interactions in COF-1 despite the polarization of the heteroatoms in the boroxine linkage groups. The attraction can be attributed to so-called charge penetration effects, which have been described frequently for interacting organic molecules[63–65] and which originate from the interpenetration of the charge clouds of non-bonded chemical entities. A more detailed discussion of charge penetration effects will be provided in section 3.4 below.

In contrast to the van der Waals and electrostatic interactions, the combined energy contribution due to Pauli repulsion and orbital rehybridization is always repulsive with a maximum of 3601 meV for the cofacial arrangement. This significantly destabilizes that structure. Another fundamental difference between $\Delta E_{orb,Pauli}$ and the other energy contributions is that $\Delta E_{orb,Pauli}$ changes significantly already for small displacements and then levels off for larger displacements. This is the primary reason, why the sum of $\Delta E_{vdW}$, $\Delta E_{elstat}$, and $\Delta E_{orb,Pauli}$, first drops with the displacement, then forms a minimum at 1.75 Å, and finally rises again for larger displacements. I.e., $\Delta E_{orb,Pauli}$ is the factor that is actually responsible for the energy minimum corresponding to a shifted rather than a cofacial arrangement of successive layers.

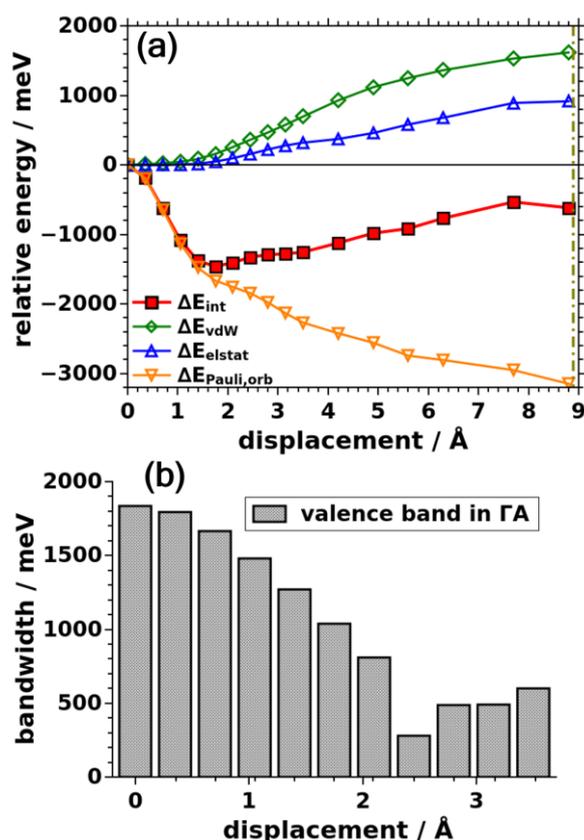

*Figure 3. Relative energies of COF-1 displaced parallel to a pore wall at a constant interlayer distance. (a) Comparison of interaction energy, vdW energy, electrostatic energy, and Pauli repulsion plus orbital rehybridization energy for the displacements. (b) Corresponding width of the valence band along a **k**-path parallel to the stacking direction for COF-1 layers as a function of the displacement. Energy values at 0.0 Å displacement: $\Delta E_{int}$=-1290 meV, $\Delta E_{int,elec}$=2257 meV, $\Delta E_{vdW}$=-3547 meV, $\Delta E_{electrostatic}$=-1344 meV, $\Delta E_{Pauli,orb}$=3601 meV, $E_{total}$=-70442.671 eV. All energies are specified per unit cell containing two COF layers. The dash-*

*dotted line indicates a shift of half the unit-cell length, beyond which the curve is simply mirrored for larger shifts.*

## 3.2 Band widths and Pauli repulsion

The crucial role of $\Delta E_{orb,Pauli}$ raises the question, why Pauli repulsion is so large for a cofacial structure. To understand that, one has to keep in mind that when occupied orbitals of two molecules overlap, bonding and antibonding linear combinations are formed, where the bonding one is stabilized less than the antibonding one is destabilized. Furthermore, the energies of the occupied bands (orbitals) enter into the expression of the total energy. Therefore, wavefunction overlap involving fully occupied orbitals results in a repulsive contribution. This effect is particularly pronounced for large energetic splittings between the bonding and antibonding states and, correspondingly, for strong electronic couplings and large bandwidths. As a consequence of orbital symmetries, this typically is the case for cofacial arrangements of π-conjugated systems. A maximum bandwidth for a vanishing displacement is, indeed, observed also here, as is shown for the case of the valence band in Figure 3b.

When the layers are shifted relative to each other, the bandwidth decreases, essentially vanishes around a displacement of 2.5 Å and then increases again, a trend that has been intensively discussed for organic semiconductors.[55,62,66–70] The shift at which the bandwidth reaches its minimum value depends on the symmetry and nodal structure of the lattice periodic functions in the Bloch states constituting the different bands. Independent of that, bandwidths and, thus, Pauli repulsion are expected to be maximized for zero-displacement for all occupied bands. Additionally, both quantities should drop for small displacements, while the shift at which the minimum is reached is band-dependent. Therefore, one cannot expect a one-to-one correlation between the valence bandwidth and Pauli repulsion. This is discussed in more detail for quinacridone and pentacene in Ref [62]. Consistently, as can be seen in Figure 3 also in COF-1 the bandwidth and total energy adopt maximum values in the cofacial case, and both values drop at very small displacement, while there is no one-to-one correlation at larger displacements.

In the case of porous materials, there is an additional aspect which goes beyond what is observed in organic semiconductors. Its origin is sketched in Figure 4, where one can see that different types of phenylene stacks, are affected differently by the displacement. Phenylene rings in stacks 2 and 3 are shifted towards open pore space, such that the associated Pauli repulsion has to drop, due to its dependence on wavefunction overlap. This is the primary reason, why $\Delta E_{Pauli,orb}$ drops over the entire range of displacements shown in Figure 3a. Conversely, the phenylene units in stack 1 experience the more "traditional" situation that a shift reduces the overlap with one entity, but increase that with another (here the boroxines). This is also the situation typically observed in organic semiconductors and relates to the discussion in the previous paragraph.

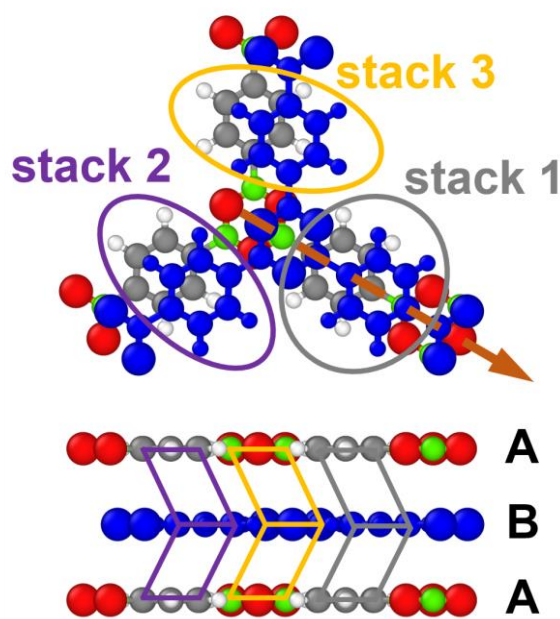

*Figure 4.* Illustration of the consequences of a shift between consecutive layers for a section of COF-1. The plot shows, how different types of phenylene stacks (denoted as stack 1, 2, and 3) are differently affected by the displacement. Here, the lattice vectors are omitted for clarity.

To more directly illustrate the stack-dependent impact of the slip on the electronic structure of COF-1, it is useful to consider the associated band structures.[32] These are shown in Figure 5 in approximately the energy range spanned by the valence band. For the cofacial arrangement displayed in Figure 5a one observes a three-fold degenerate valence band, which is essentially flat for **k**-vectors within the plane of the layers (with several additional bands between 0.0 and 0.35 eV). Such flat bands are actually expected for systems with three-arm cores and three-fold symmetry.[71] The situation changes significantly in the ΓA direction, where the effective bandwidth of the valence band amounts to ~1.8 eV, when considering the backfolding of the band due to the two layers in the unit cell. The densities of states projected onto the individual sub-units of the COF, namely the phenylene stacks 1, 2, and 3, and the boroxine stack (the PDOSs shown in Figure 5c), show that for zero displacement the degenerate valence bands are dominated by equal contributions of states on the three phenylene linkers. The boroxine-related states are found at slightly more negative energies.

For the minimum energy conformation (with a displacement of 1.75 Å along the pore wall), the situation changes fundamentally (see Figures 5b and d): The degeneracy of the valence bands is lifted and the topmost band is derived only from phenylene stack 1. This is apparent from the plot of the PDOSs in Figure 5d and from the electron density corresponding to the highest occupied eigenstate in Figure 5e. The width of that band is reduced to ~1 eV. The bandwidth reduction is even more pronounced for most of the lower-lying bands associated with stacks 2 and 3 (consistent with the diminishing overlap of the respective phenylene units).

These changes in the band structure have a distinct impact on charge-transport related properties of the serrated system: The interlayer electronic coupling is reduced in line with the bandwidth by a factor of ~1.8. This is a consequence of the coupling being directly proportional to the bandwidth for simple, cosine shaped bands like the valence band of COF-1 (as can be inferred, for example, from a tight-binding description; see also dashed green line in Figure 5b). Additionally, the effective mass for transport in π-stacking direction more than doubles from $0.61 \cdot m_0$ in the cofacial case to $1.29 \cdot m_0$ for the minimum energy structure (with $m_0$ corresponding to the free electron mass). On more technical grounds, we note that the above trends prevail, when employing the hybrid HSE06[53,54] functional instead of the PBE[47,48] functional, as shown in the Supporting Information.

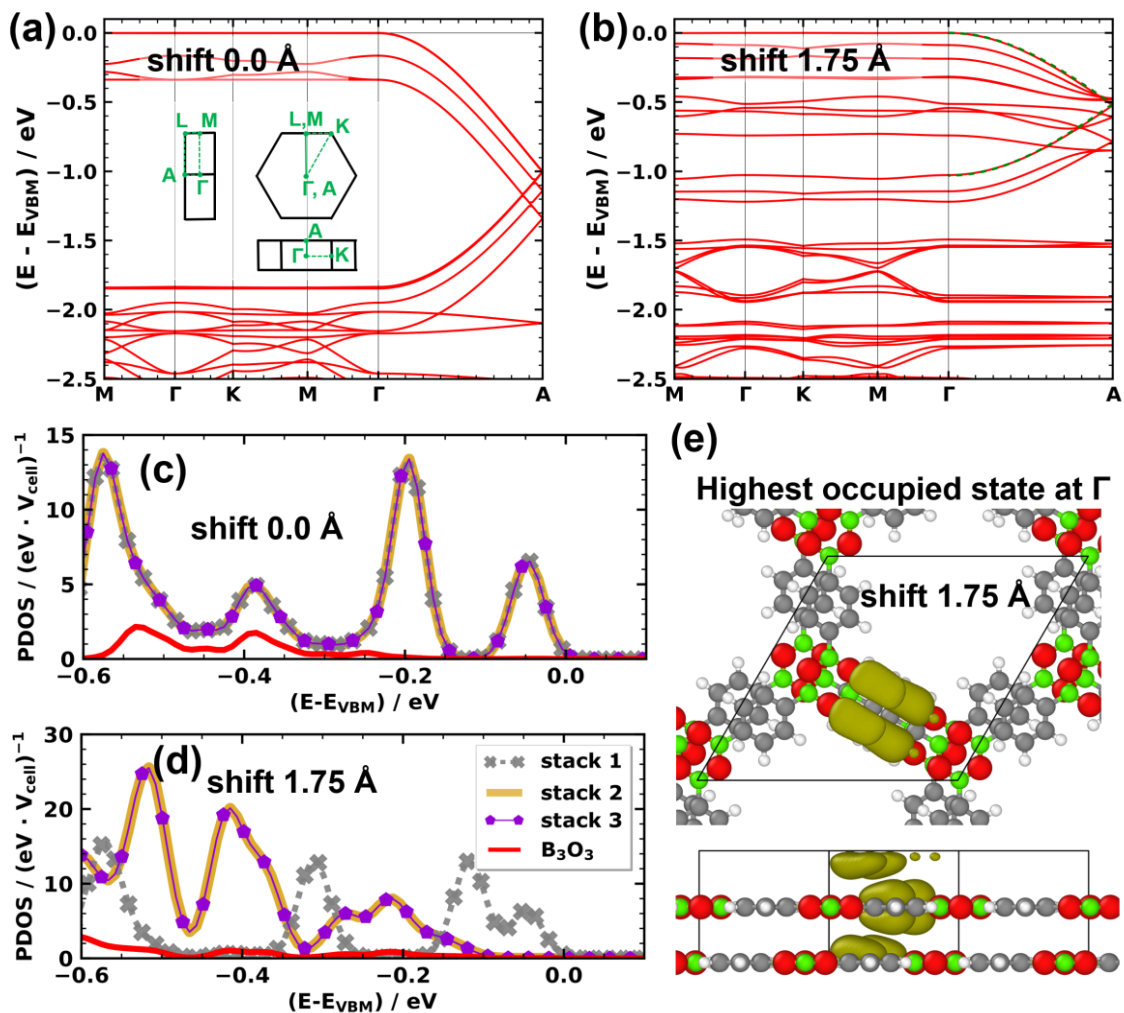

*Figure 5.* Electronic band structure ((a) and (b)), projected density of states ((c) and (d)), and eigenstate density ((e)) of COF-1 for the cofacial arrangement ((a) and (c)) and for the minimum energy, serrated structure ((b), (d), and (e)) with successive layers shift by 1.75 Å along the direction of a pore walls. All quantities have been calculated for a structure with the interlayer distance fixed to 3.328 Å. The densities of states are projected onto the phenylene stacks 1, 2, and 3 (grey, purple, dark yellow, see Figure 4) and the boroxine-based stack (red). The energy range in the PDOS plots is reduced compared to the band-structure plots. The dashed green line in panel (c) is the result of a fit of a simple 1D tight-binding model and serves to illustrate the structure of the valence band. In (e) an isodensity plot of the electron density of associated with the valence band state at the $\Gamma$ point is shown. Color code of the atoms: C ... grey, H ... white, B ... green, O ... red;

## 3.3 Optimized Interlayer Stacking Distance

While the above considerations provide relevant fundamental insights concerning the consequences of a displacement of neighboring 2D COF layers, they disregard the fact that an increased interlayer repulsion/attraction triggers an increase/decrease of the interlayer stacking distance. Therefore, as a next step, we studied, how the situation changes when the stacking distance between consecutive COF layers is optimized for each displacement. As this optimization is computationally rather demanding (see Methods section and Supporting Information), we here focus on shifts close to the energetic minimum (found above for a constant interlayer distance). In passing we note that now the evolution of the total energy no longer overlaps directly with that of the interaction energy, as due to the varying interlayer distance the calculations for the sub-systems A and B change as a function of the shift. Still, evolutions of the total and the interaction energies evolve essentially in parallel, as shown in the Supporting Information.

The data on the interaction energy in Figure 6 reveal that the energy minimum for shifts parallel to the pore wall is still found around 1.75 Å, just like for the constant interlayer stacking distance. This can be rationalized by the observation that the calculated optimum interlayer distance (3.350 Å) considered in Figure 6a is very close to the experimental value (3.328 Å) used for obtaining the trends in Figure 3a (see also dash-dotted line in Figure 6b). This is insofar rather surprising, as now the evolutions of all contributions to the interaction energy with displacement are fundamentally different from the situation discussed above. This goes far beyond a mere reduction of the magnitudes of the variations, which one would expect because of the additional "degree of flexibility" of the system.

Now, the van der Waals attraction no longer decreases continuously with displacement as in Figure 3. Rather, the vdW energy becomes more negative up to a displacement of 1.75 Å and rises only afterwards. One can rationalize this behavior by a pronounced decrease of the interlayer stacking distance for small displacements relative to the cofacial case (Figure 6b). This overcompensates the consequences of the decreased lateral overlap of the layers, as can be quantified by weighted distance histograms, which are contained in the Supporting Information. This overcompensation vanishes for displacements beyond 1.75 Å, where the change in stacking distance is less pronounced and where also larger portions of the COF layers come to lie above the (empty) pores of neighboring layers. In passing we note that, the decrease in interlayer distance with displacement roughly follows the evolution of $\Delta E_{Pauli,orb}$ in Figure 3a. This is, in fact, sensible, considering that (i) Pauli repulsion is by far the largest contribution in Figure 3a and that (ii) it depends on wavefunction overlap and should, thus, display a particularly pronounced distance dependence.

In line with this reasoning, the decrease in interlayer distance with layer displacement essentially inverts the situation for Pauli repulsion: $\Delta E_{Pauli,orb}$ remains strongly positive (i.e., repulsive) for all distances, but now, the decrease in wavefunction overlap due to the shape of the orbitals (vide supra) for shifted systems is overcompensated by an increase of the repulsion due to the larger wavefunction overlap at smaller distances. Only for larger displacements $\Delta E_{Pauli,orb}$ drops again because of an increasing fraction of the COF coming to lie above the pores of neighboring layers. The consequences of the above-described compensation effects are visible also in the evolution of the valence bandwidth (see Figure 6c), which now displays a pronounced decrease only for shifts beyond the equilibrium displacement of 1.75 Å. Therefore, the cofacial arrangement also loses part of its advantage over the energetic minimum structure as far as bandwidths (and the resulting interlayer electronic coupling) and effective masses are concerned. The bandwidth only decreases from 1.15 eV to 1.00 eV. For the effective mass, the effect is still a bit larger with an increase from $0.85 \cdot m_0$ in the cofacial case to $1.31 \cdot m_0$ for a layer displacement of 1.75 Å parallel to the pore wall.

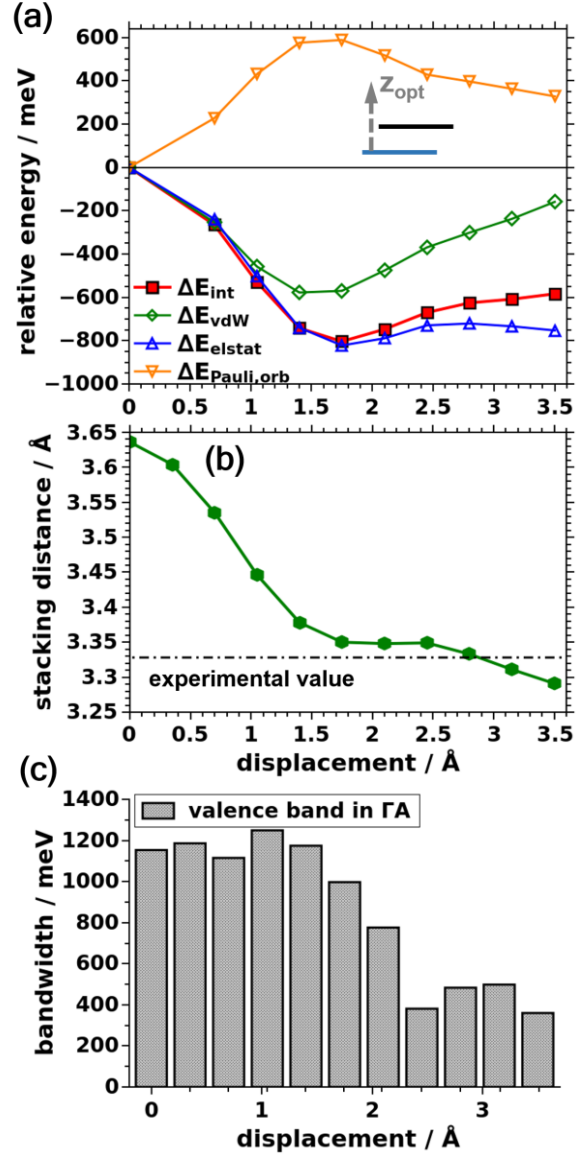

*Figure 6. Relative energies, stacking distance, and valence bandwidth of COF-1 displaced parallel to one of the pore walls. Here, the stacking distance has been optimized for each displacement. (a) Comparison of interaction energy, vdW energy, electrostatic energy, and Pauli plus orbital rehybridization energy; (b) optimized stacking distance between consecutive COF-1 layers; (c) width of the valence band along a **k**-vector parallel to the stacking direction of COF-1 layers. Energy values at 0.0 Å displacement: $\Delta E_{int}$=-1957 meV, $\Delta E_{int,elec}$=812 meV, $\Delta E_{vdW}$=-2769 meV, $\Delta E_{elstat}$=-385 meV, $\Delta E_{Pauli,orb}$=1197 meV, $E_{total}$=-70443.275 eV. All energies are specified per unit cell containing two COF layers.*

Interestingly, also the evolution of the (attractive) electrostatic interaction is largely inverted compared to the situation for fixed interlayer distances and the magnitude of the changes is significantly increased. For small displacements, this evolution is again caused by the significantly decreasing interlayer distances. This amplifies the attractive charge-penetration effects for shifted layers. Only at larger displacements the diminishing overlap of significant parts of the COF layers becomes again the dominant factor and results in a reduction of the electrostatic interaction. As a consequence, when allowing the interlayer distance to relax for displaced structures, electrostatic interactions do favor the serrated configuration. This is, however, again not caused by a "conventional" electrostatic repulsion due to a high octupole moment of the $B_3O_3$ rings, as suggested by chemical intuition. Rather, the evolution is primarily a consequence of the changing inter-layer distance with small distances resulting in an increased electrostatic attraction. This calls for a more in-depth investigation of the role played by the $B_3O_3$ ring, which will be provided in the next section.

## 3.4 Attractive Electrostatic Energy and Charge Penetration Effect

In all situations encountered so far, the electrostatic interactions between the layers were attractive (i.e., the electrostatic energy was negative). As mentioned before, this is commonly attributed to charge penetration, an effect that has been discussed extensively for interacting organic materials and organic semiconductor crystals.[63–65] Conceptually, this effect describes that due to the interpenetration of charge/electron clouds the shielding of the positively charged nuclei is reduced and the attractive electron-nuclei interaction increases.

As indicated above, for systems like COF-1 one would still expect that the significantly different electronegativities of the B and O atoms in the central $B_3O_3$ rings would result in sizable octupole moments resulting in a repulsion of cofacial $B_3O_3$ rings. To disentangle the roles of the phenylenes and the boroxine-based linkages, we split the individual COF-1 layers into two model systems consisting either of benzene or boroxine ($B_3O_3H_3$) rings (i.e., into an only weakly polar and a highly polar unit; see insets in Figure 7a). The dangling bonds of the model systems were saturated by H atoms and the rings were arranged at exactly the same positions they adopt in COF-1 with optimized interlayer distances. Then, $\Delta E_{elstat}$ was calculated separately for each model system as a function of the shift of consecutive layers. The resulting energy evolution is shown in Figure 7a for constant interlayer stacking distances and in Figure 7c the situation for optimized distances can be found. For both cases, when considering only the benzene molecules, $\Delta E_{elstat}$ is indeed primarily determined by charge penetration effects and is clearly attractive for all considered situations in analogy to the situation observed above for the full COF. Interestingly, when considering only the boroxine units, for small displacements the electrostatic interactions become repulsive, in line with the sizable octupole moments of the molecules. This is insofar relevant, as the electrostatic interaction between the layers in COF-1, in a first approximation, can be regarded as a superposition of the (largely independent) interactions of the phenylene subsystem and the boroxine subsystems, as shown in Figures 7b,d (again for constant as well as for optimized interlayer stacking distances). This implies that there is, indeed, electrostatic repulsion between the boroxines at small displacements also in the actual COF-1. This repulsion is, however, overcompensated by the attractive interactions of the phenylenes, resulting in the final trends observed in Figures 3 and 6. This overcompensation is clearly visible for both cases, comprising constant and optimized stacking distances. Interestingly, for constant interlayer stacking distance, this overcompensation is even strong enough that the cofacial arrangement of neighboring sheets, which is electrostatically unfavorable when considering only the boroxine units, becomes favorable for the full COF-1 layers.

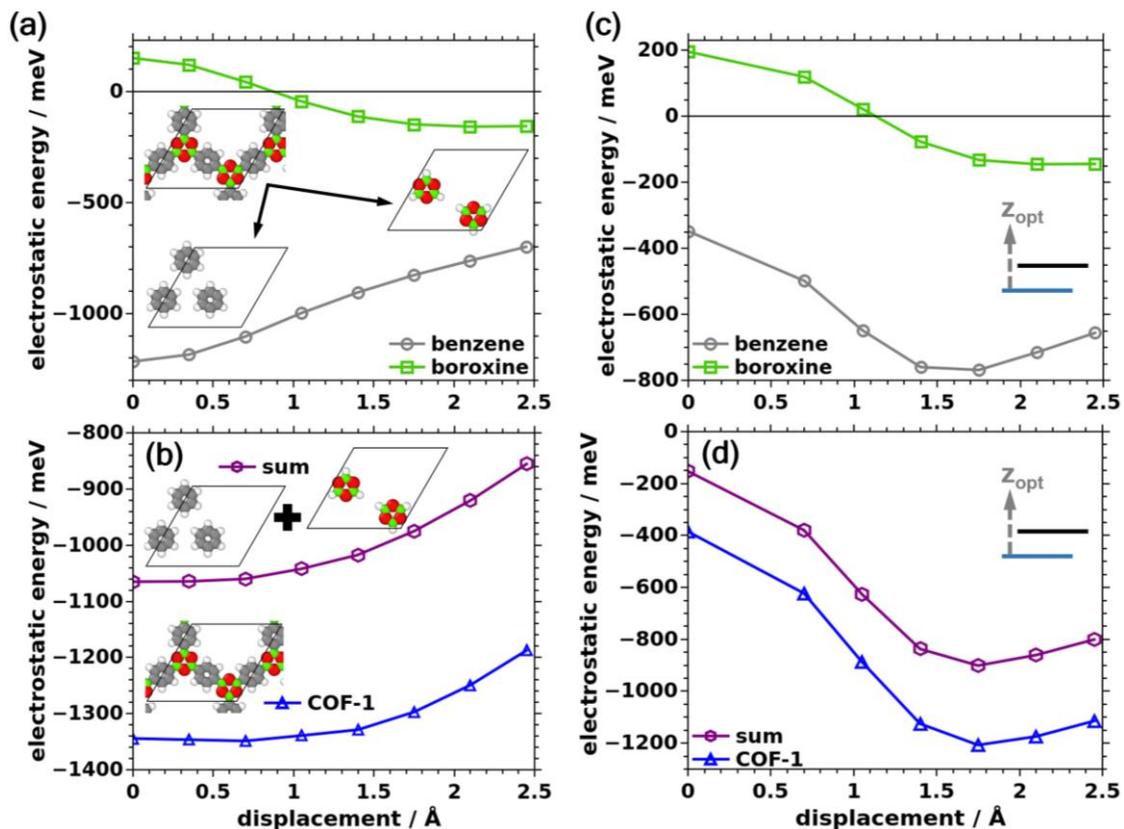

*Figure 7. Panels (a) and (c) show the electrostatic energy contributions of the COF-1 sub-systems (boroxine unit and benzene) extracted from COF-1 with constant (a) and optimized (c) interlayer stacking distances. The nature of the sub-systems is indicated by the insets in panel (c). In panels (b) and (d) the sum of the electrostatic energies stemming from the two sub-systems is shown by the purple line and symbols and the electrostatic energy of COF-1 is given in blue. Panel (b) shows the results for constant and panel (d) for optimized interlayer stacking distances. Note that in contrast to Figures 3 and 6, the electrostatic energies are not given relative to the situation for zero displacement. Rather, their absolute values are plotted, as when separately considering the boroxines, the electrostatic energy changes sign as a function of the displacement. All energies are specified per unit cell containing two COF layers*

## 3.5 Additional Layered COFs

In the following, we will briefly address to what extent the above observations are specific to COF-1 or whether they can be considered to be of more general nature. Therefore, we considered additional 2D COFs with π-systems of different size and nature and also with different pore topologies. These systems comprise COF-5, as a more extended analogue to COF-1, and the porphyrin- and hexabenzocoronene-based COFs Por-COF,[44] and HBC-COF.[39] Their structures are shown in Figure 8. For Por-COF we consider two versions of that system, one consisting of Zn-metallated porphyrin (Zn) and one without a metal incorporated in the center of the molecule (NH).

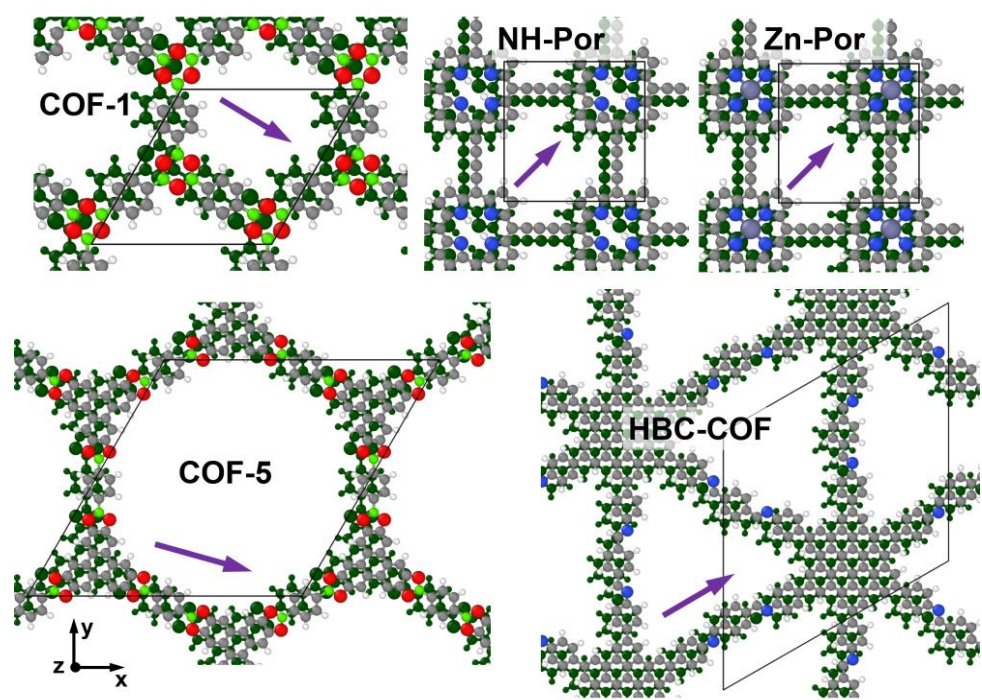

*Figure 8.* Optimized structures of the considered COFs: COF-1, COF-5, HBC-COF, NH-Por, Zn-Por. For each COF the respective unit cell containing two original layers is shown. The atoms of the top layer are colored according to their chemical nature (C ... grey, H ... white, B ... green, O ... red, N ... blue, Zn ... violet). The atoms in the displaced bottom layer are all plotted in dark green. The layer displacements of all COFs are indicated by purple arrows. Interestingly, the resulting packing motif of the aromatic systems of COF-1, COF-5, and HBC-COF, resemble that of graphite.[36]

When performing a full geometry optimization (for details see methods section), all of these COFs adopt a serrated structure (see Figure 8). This is consistent with the expectation from literature that apart from a few exceptions such shifted AA-stackings are the preferred stacking motif.[36] The obtained lateral displacement vectors, $\Delta v_{xy}$, between consecutive layers are reported in Table 2 and indicated by purple arrows in Figure 8. Considering the absolute values of these vectors, we find that they are of similar magnitude for all COFs. Furthermore, from the interlayer stacking distances reported in Table 2 one can see that the energetically favorable serrated layer arrangements exhibit significantly smaller stacking distances compared to the cofacial arrangement.

*Table 1.* *Structural parameters (displacement vectors, stacking distances) of the considered COFs. $\Delta v_{xy}$ … displacement vector for the second layer in the unit cell between cofacial and optimized structure, z … stacking distance (given for the cofacial and the optimal arrangement of the COF).*

|  |  | COF-1 | COF-5 | HBC-COF | NH-Por | Zn-Por |
|---|---|---|---|---|---|---|
| $\Delta v_{xy}$ / Å | | (1.50,-0.86) | (1.49, -0.46) | (0.00,1.60) | (1.18, 1.18) | (1.24, 1.24) |
| z / Å | cofacial | 3.62 | 3.59 | 3.62 | 3.55 | 3.56 |
|  | optimal | 3.36 | 3.39 | 3.43 | 3.33 | 3.32 |

To understand, whether similar driving forces as in COF-1 are responsible for the serrated AA-stacking, Figure 9 compares the changes in interaction, van der Waals, electrostatic and Pauli repulsion plus orbital rehybridization energies between cofacial structures with optimized interlayer distances and optimized structures for all considered COFs. The absolute values of the energies are contained in the Supporting Information. All COFs display the same behavior that has been discussed in section 3.3 for COF-1: The displaced structure is stabilized by an increased van der Waals and Coulomb interaction while the Pauli repulsion increases consistent with the observation in Figure 6. Considering that also the relative magnitudes of the individual contributions are reasonably similar in all systems, one can confidently attribute these changes to a superposition of the effects due to the displacement and the concomitantly decreased interlayer stacking distance in the serrated structure (see Table 2) in analogy to the discussion in section 3.3. This is shown explicitly for COF-5 in the Supporting Information, where equivalent trends as in Figure 6 are shown for the interaction energy and the individual contributions (van der Waals, Coulomb, and Pauli repulsion plus orbital rehybridization).

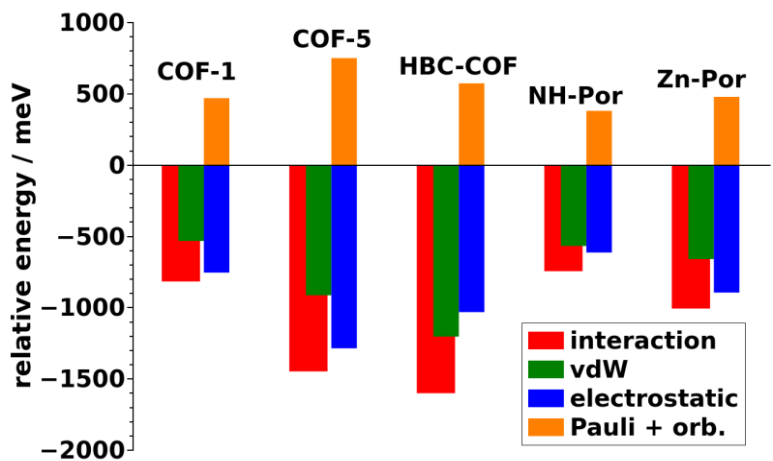

*Figure 9.* Energy differences between the cofacial and the optimized structures of the COF-1, COF-5, HBC-COF, NH-Por, Zn-Por. Differences in the total interaction energy are compared to the contributions, van der Waals, electrostatic, and Pauli repulsion plus orbital rehybridization interactions. Minor differences in the values for COF-1 in Figures 9 and 6 are due to the fact that for Figure 6 we relied on the experimental lateral unit cell parameters, while here, for the sake of comparability with the other COFs, all unit-cell parameters were optimized.

## 4. CONCLUSIONS

Based on the prototypical example of COF-1, we identified the driving forces that result in shifted (serrated) AA-stacking arrangements to be energetically favorable. A quantitative assessment of the individual interactions (dispersion, electrostatic, Pauli repulsion plus orbital rehybridization) determining the evolution of the total energy is provided employing dispersion-corrected density-functional theory. For both, fixed and variable interlayer distances the actual equilibrium structure is determined by a subtle interplay between van der Waals and Coulomb attraction as well as Pauli repulsion. How these different interactions play out is, however, fundamentally different in the two cases: In the fixed distance case, van der Waals and Coulomb attractions favor a cofacial arrangement as displacement of consecutive layers increases the net inter-atomic distances and, thus, diminishes the attraction due to dispersion forces and charge penetration effects. Conversely, Pauli repulsion favors displaced structures, which can be correlated at least in part with a decrease of the width of the valence band at small displacements. The latter also results in clearly deteriorated transport parameters for the minimum structure with a decrease of the electronic coupling and an increase of the effective mass of the holes in the valence band by a factor of roughly two.

The role of the different contributions is essentially inverted when optimizing the interlayer distance. The reason for that is that the effects resulting from a decreased geometrical overlap in the displaced structures are compensated by the impact of the concomitantly observed decreased interlayer distance. Considering the fundamentally different origins of Coulomb, Pauli, and van der Waals interactions, these compensation effects play out differently. Nevertheless, the optimum displacement is essentially the same for both situations, comprising a shift of 1.75 Å parallel to one of the pore walls.

The above results, at first glance, suggest that "conventional" electrostatic repulsion between the highly polarized boroxine units, which has previously been held responsible for the serrated structure of COF-1,[37,43] actually does not play a role. This would contradict chemical intuition and a more in-depth analysis of the contribution of the boroxines to the electrostatic energy reveals that they, indeed, cause electrostatic repulsion for a cofacial arrangement of the layers. This repulsion is, however, overcompensated by the attractive interactions of the phenylenes due to charge penetration. A similar behavior as in COF-1 is found for a series of additional COFs with differently sized π-systems and differently shaped pores.

Overall, our findings imply that the structure of layered COFs is to a significant extent determined by effects that are of quantum-mechanical nature (like Pauli repulsion) and that depend on the symmetry and nodal structure of extended electronic states. As a consequence, approaches relying, for example, on classical force fields should not be able to provide a qualitatively correct description of the relevant physics. Even more importantly, our findings suggest that for obtaining layered COFs with cofacial stacking arrangements (as would be ideal for charge-transport applications), one cannot rely on the self-assembly of the individual layers. Also, merely reducing the classical electrostatic repulsion between layers by eliminating polar functionalities will not resolve the issue. Instead, one has to introduce additional driving forces (e.g., due to steric effects) in order to realize the desired stacking, tipping the otherwise unfavorable balance between van der Waals, Coulomb, and Pauli interactions.

AUTHOR INFORMATION

**ORCID**

Christian Winkler     0000-0002-7463-6840

Egbert Zojer          0000-0002-6502-1721

**Corresponding Author**

egbert.zojer@tugraz.at

**Author Contributions**

E. Zojer and C. Winkler conceptualized the present study of the stacking arrangements of covalent organic frameworks and the decomposition into the determining interactions. C. Winkler performed all calculations and did a primary analysis of the data. C. Winkler also wrote a first version of the manuscript and prepared all figures. The present version of the manuscript was revised by E. Zojer and C. Winkler. The entire project was supervised by E. Zojer.

**Funding Sources**

TU Graz Lead Project "Porous Materials at Work" (LP-03).

**Conflicts of Interest**

There are no conflicts to declare.

ACKNOWLEDGMENT

The work has been financially supported by the TU Graz Lead Project "Porous Materials at Work" (LP-03). The computational results have been in part achieved using the Vienna Scientific Cluster (VSC3).

REFERENCES

1  P. Kuhn, M. Antonietti and A. Thomas, *Angew. Chemie - Int. Ed.*, 2008, **47**, 3450–3453.

2  A. P. Côté, A. I. Benin, N. W. Ockwig, M. O'Keeffe, A. J. Matzger and O. M. Yaghi, *Science*, 2005, **310**, 1166–1170.

3  H. M. El-Kaderi, J. R. Hunt, J. L. Mendoza-Cortés, A. P. Côté, R. E. Taylor, M. O'Keeffe and O. M. Yaghi, *Science*, 2007, **316**, 268–272.

4  R. W. Tilford, W. R. Gemmill, H. C. Zur Loye and J. J. Lavigne, *Chem. Mater.*, 2006, **18**, 5296–5301.

5  F. J. Uribe-Romo, J. R. Hunt, H. Furukawa, C. Klöck, M. O'Keeffe and O. M. Yaghi, *J. Am. Chem. Soc.*, 2009, **131**, 4570–4571.

6  A. P. Côté, H. M. El-Kaderi, H. Furukawa, J. R. Hunt and O. M. Yaghi, *J. Am. Chem. Soc.*, 2007, **129**, 12914–12915.

7  X. Chen, K. Geng, R. Liu, K. T. Tan, Y. Gong, Z. Li, S. Tao, Q. Jiang and D. Jiang, *Angew. Chemie - Int. Ed.*, 2020, **59**, 5050–5091.

8  H. Furukawa and O. M. Yaghi, *J. Am. Chem. Soc.*, 2009, **131**, 8875–8883.

9  R. W. Tilford, S. J. Mugavero, P. J. Pellechia and J. J. Lavigne, *Adv. Mater.*, 2008, **20**, 2741–2746.

10  Y. Yang, M. Faheem, L. Wang, Q. Meng, H. Sha, N. Yang, Y. Yuan and G. Zhu, *ACS Cent. Sci.*, 2018, **4**, 748–754.

11  Z. Kang, Y. Peng, Y. Qian, D. Yuan, M. A. Addicoat, T. Heine, Z. Hu, L. Tee, Z. Guo and D. Zhao, *Chem. Mater.*, 2016, **28**, 1277–1285.

12  H. L. Qian, C. X. Yang and X. P. Yan, *Nat. Commun.*, 2016, **7**, 1–7.

13  L. A. Baldwin, J. W. Crowe, D. A. Pyles and P. L. McGrier, *J. Am. Chem. Soc.*, 2016, **138**, 15134–15137.

14  K. Dey, M. Pal, K. C. Rout, S. S. Kunjattu, A. Das, R. Mukherjee, U. K. Kharul and R. Banerjee, *J. Am. Chem. Soc.*, 2017, **139**, 13083–13091.

15  H. Li, Q. Pan, Y. Ma, X. Guan, M. Xue, Q. Fang, Y. Yan, V. Valtchev and S. Qiu, *J. Am. Chem. Soc.*, 2016, **138**, 14783–14788.

16  S. Lin, C. S. Diercks, Y. B. Zhang, N. Kornienko, E. M. Nichols, Y. Zhao, A. R. Paris, D. Kim, P. Yang, O. M. Yaghi and C. J. Chang, *Science*, 2015, **349**, 1208–1213.

17  Q. Sun, B. Aguila, J. Perman, N. Nguyen and S. Ma, *J. Am. Chem. Soc.*, 2016, **138**, 15790–15796.

18  S. Wang, Q. Wang, P. Shao, Y. Han, X. Gao, L. Ma, S. Yuan, X. Ma, J. Zhou, X. Feng and B. Wang, *J. Am. Chem. Soc.*, 2017, **139**, 4258–4261.

SUPPORTING INFORMATION for

# *Nature of the interactions determining the stacking motif of covalent organic frameworks*

*Christian Winkler[1] and Egbert Zojer[1,]\**

[1] Institute of Solid State Physics, NAWI Graz, Graz University of Technology, Petersgasse 16, 8010 Graz, Austria

# 1. Methodological details

Please note that all input and the most important output files of all calculations can be found on the NOMAD database at: https://dx.doi.org/10.17172/NOMAD/2021.02.15-1

Thus, the details on all parameters and settings that have been employed during the calculations can be found there.

## 1.1. Overview of basis functions used in FHI-Aims

*Table S1. Basis functions that have been used for all calculations performed with FHI-AIMS. The abbreviations read as follows: H(nl,z), where H refers to hydrogen-like basis functions, n is the main quantum number, l denotes the angular momentum quantum number, and z denotes an effective nuclear charge which scales the radial function in the defining Coulomb potential.[1]*

|  | H | C | B | O | N | Zn |
|---|---|---|---|---|---|---|
| Minimal | 1s | [He]+2s2p | [He]+2s2p | [He]+2s2p | [He]+2s2p | [Ar]+4s3p3d |
| Tier 1 | H(2s,2.1) H(2p,3.5) | H(2p,1.7) H(3d,6) H(2s,4.9) | H(2p,1.4) H(3d,4.8) H(2s,4) | H(2p,1.8) H(3d,7.6) H(3s,6.4) | H(2p,1.8) H(3d,6.8) H(3s,5.8) | H(2p,1.7) H(3s,2.9) H(4p,5.4) H(4f,7.8) H(3d,4.5) |
| Tier 2 | H(1s,0.85) H(2p,3.7) H(2s,1.2) H(3d,7) | H(3p,5.2) H(3s,4.3) H(3d,6.2) H(4f,9.8) H(5g,14.4) | H(4f,7.8) H(3p,4.2) H(3s,3.3) H(5g,11.2) H(3d,5.4) | H(3p,6.2) H(3d,5.6) H(1s,0.75) H(4f,11.6) H(5g,17.6) | H(3p,5.8) H(1s,0.8) H(3d,4.9) |  |

## 1.2. Determination and optimization of individual COF-structures

Let us note some general aspects before we describe the details how the individual structures of the considered COFs have been obtained. First of all, we grouped the systems into two categories: category (A) comprises COF-1, for which reliable experimental lattice parameters are available[2] and category (B) which comprises COF-5, Zn-Por, NH-Por, and HBC-COF. This categorization was employed, as for several of the materials in category (B) no complete set of experimental lattice parameters is available. Therefore, for these materials all unit cell parameters (in-plane parameters and cell-heights) had to be determined. In the following we will first describe how the unit cells of the materials falling under category (B) were constructed. Further, we will discuss category (A), i.e. COF-1 with the experimental lattice parameters. All these considerations will be for the initial cofacial interlayer arrangements of the systems. Then we will describe how the shifted (or displaced) structures have been obtained.

### 1.2.1 Construction of the unit cells for COF-5, NH-Por, Zn-Por, and HBC-COF

The unit cell parameters of these systems were obtained following a two-step procedure. In a first step **(1)**, the in-plane lattice parameters were calculated and then in a next step **(2)**, based on these in-plane parameters, the optimal stacking distance for the cofacial arrangement was evaluated. The main reason for this stepwise procedure was to obtain a cofacial arrangement, avoiding shifts of consecutive layers in a full geometry optimization. These would be energetically favorable, as discussed in the main manuscript. In detail, we performed the following steps:

**(1)** The optimal in plane lattice parameters for the individual COFs were evaluated by considering a COF monolayer (4x4x1 k-point grid, total energy converged within less than 1 meV, individual layers were decoupled quantum-mechanically by a vacuum of 40 Å and electrostatically by using a dipole correction) and gradually shrinking the lateral unit cell size while keeping the initial symmetry. For each unit cell size, all atomic positions were relaxed[‡] and the total energy was calculated. These data were fitted with a Birch-Murnaghan equation of state[3] to obtain the equilibrium in-plane lattice parameters.

**(2)** The obtained in-plane unit cell parameters and the atomic positions of the relaxed monolayer are then used for constructing the bulk structure of each COF. These unit cells contain two consecutive layers (A and B) in stacking direction, where these layers are cofacially stacked at an initial distance of 4 Å. This stacking distance then was varied within a range of ±0.75 Å in steps of 0.25 Å. For COF-5, Zn-Por, and NH-Por single point calculations were performed to get the total energy of the system for each stacking distance. For the only non-planar COF considered here, HBC-COF, the atomic positions were relaxed at each stacking distance and from these relaxations the total energy was obtained. During these relaxations, the planar hexabenzocoronene-core was fixed with respect to in-plane displacements. This relaxation step was necessary, as especially the phenylene groups of individual COF-layers can twist. Now, based on the total energy as a function of displacement, we identified the minimum and calculated additional data points (single-point calculations for COF-5, Zn-Por, and NH-Por and geometry relaxations for HBC-COF) around this minimum with ±0.125 Å variation. All data points were then fitted with a Birch-Murnaghan equation of state and the minimum of that fit was used as the optimal stacking distance of layers A and B.

Finally, the atomic positions of the COFs in the obtained unit cells were relaxed. In order to avoid interlayer shifts, the in-plane positions of 2 atoms contained in the planar core of the individual systems were fixed. These planar cores are the triphenylene units in COF-5, the porphyrins in Zn-Por and NH-Por and the hexabenzocoronene in HBC-COF.

Note that this procedure was employed to construct the unit cell for the cofacial arrangement of the individual COFs. These structures and unit cells then serve as the starting points for finding the systems with the lowest total energy. Details how this was done can be found below in section 1.2.4..

### 1.2.2 Construction of the unit cell for COF-1

COF-1 is the system at the heart of our investigations. For this system the unit cell has been constructed in two ways (i) and (ii). Type (i) was constructed by employing the experimental lattice parameters (in-plane: a=b=15.420 Å and stacking distance z=3.328 Å) reported in literature.[2] Type (ii) relied on relaxed unit cell parameters following the relaxation procedure outlined in section 1.2.1.[§] Essentially all data for COF-1 in the main manuscript were obtained for the system built from the experimental lattice parameters. Therefore, we start by describing the optimization procedure for this system.

---

[‡] For all geometry relaxations the convergence criterion was set to 0.01 eV/Å. This means that all atomic positions were relaxed until the largest force component on any of the atoms was below this value.

For type (i) the bulk structure of COF-1 was constructed by first optimizing the positions of all atoms in isolated monolayers (setting the unit cell height perpendicular to the layers to 40.0 Å and employing a 4x4x1 k-grid). Then these layers were stacked at the experimental interlayer distance[2] of 3.328 Å with the unit cell of COF-1 containing two layers in stacking direction (layers A and B, see Figure 1 of main manuscript). This allows displacing these layers along directions parallel to the xy-plane.

### 1.2.3 Stacking distances of COF-1 at cofacial and shifted layer arrangements

To investigate the energies as a function of interlayer displacement for constant stacking distance, one can simply displace consecutive layers along the respective direction and then calculate the energies. When considering also changes of the stacking distance one has to optimize this distance at each displacement. This was done in the following way:

One of the two layers in the unit cell was displaced along the considered shift direction, while the stacking distance was kept at the initial value of 3.328 Å. This stacking distance then was varied within a range of ±0.50 Å in steps of 0.25 Å for the shifted arrangement. The total energies of these structures were calculated in single-point calculations. Based on the obtained energies the minimum was identified and additional data points were calculated around this minimum with ±0.125 Å variation. Then the total energies as a function of the stacking distance were fitted using a Birch-Murnaghan equation of state and the minimum of that fit determines the optimal stacking distance. Here, the positions of the atoms within individual layers correspond to those when relaxing isolated monolayers. That this has hardly any impact on the results will be shown in sections 2.1 and 2.2.

### 1.2.4 Finding the optimal structures of COF-5, NH-Por, Zn-Por, HBC-COF, and COF-1

To find the layer arrangement with the lowest energy for all the COFs considered in this work, we performed geometry relaxations as follows: The obtained cofacial structures were used as a starting point with the two layers in the unit cell somewhat shifted relative to each other to avoid starting the geometry relaxation from a saddle point. In the optimizations then all atomic positions were allowed to relax together with the unit cell vector in stacking direction.

For COF-1, both systems, the one with the unit cell constructed from the experimental lattice parameters and the one possessing optimized in-plane parameters together with the optimized stacking distance for the cofacial arrangement were considered. In the main manuscript, in section "3.5 Additional Layered COFs" we include the data for the optimized parameters to stay consistent with the data for COF-5, NH-Por, Zn-Por, and HBC-COF reported in that section of the main manuscript. Here, in section 2.2 of this Supporting Information we compare the absolute energies (including the energy contributions) obtained for the two unit cells ((i) and (ii)) of COF-1. There one can see that especially the differences in these energies are rather small. Furthermore, the trends are in excellent agreement no matter whether unit cell (i) or (ii) are considered for COF-1, which suggests that the exact details on how the geometry is obtained is of only minor relevance.

---

[§] It was found that for COF-1 optimizing the in-plane lattice parameters results in smaller lattice constants of a=b=15.126 Å but does not change the trends and the relative ratios of the effects studied – see section 2.1.. These optimized in-plane lattice parameters are in good agreement with previous computational studies.[10]

## 2. Additional data

### 2.1. Relative energies of COF-1 in a fully optimized unit cell

Figure S1 shows the evolution of the relative energies as a function of the displacement parallel to a pore wall for COF-1 with unit cell parameters taken from the optimized cell (rather than from the experimental one).

For each displacement the optimal stacking distance was determined using the procedure outlined in section 1.2.3. Now, considering the evolution of the energies in Figure S1 and comparing them to Figure 3 of the main manuscript one observes that the data agree. Only minor numerical differences can be seen. Thus, one can conclude that optimizing in the in-plane lattice parameters does not impact the qualitative behavior, i.e. the trends, of the relative energies of the system.

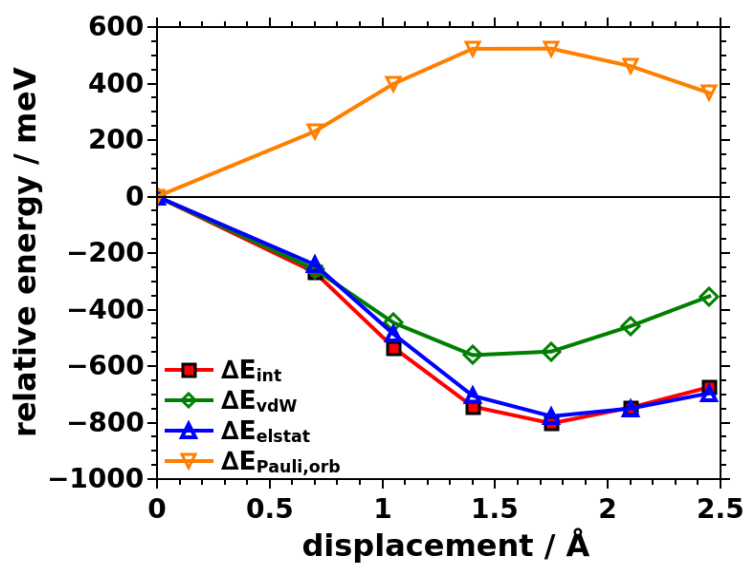

*Figure S1. Relative energies of COF-1 with optimized in-plane lattice parameters. Consecutive layers are shifted along a direction parallel to a pore wall and the stacking distance is relaxed at each displacement. (a) Comparison of total energy to the (electronic) interaction energy and the vdW energy (b) Decomposed terms of the electronic interaction energy. The energy values for 0.0 Å displacement are reported in Table S2.*

### 2.2. Comparison of absolute energies of COF-1 for cofacial and optimal arrangement

The absolute energies for a cofacial and the optimal arrangement of COF-1 layers have been obtained for unit cells (i) and (ii). Here (i) is the system with the experimental in-plane lattice parameters and (ii) denotes the system with the relaxed in-plane parameters. The obtained data are reported in Table S2.

The interlayer stacking distances show hardly any difference when comparing COF-1 with the experimental and the optimized in-plane lattice parameters. When considering the total and the interaction energies, we find that the system with the relaxed in-plane lattice parameters yields lower values, i.e. it is more stable. Dealing with the individual contributions to the interaction energy, a similar behavior can be found for the attractive vdW and electrostatic energies, the relaxed system consistently yields lower values (larger magnitudes, more negative). This trend is also found for the repulsive term comprising Pauli repulsion with orbital rehybridization. The effect, however, is different, as here the relaxed system is slightly destabilized by this energy contribution.

More important, however, is the comparison of the changes in these energies when comparing the optimal layer arrangement to the cofacial one. These values can be found in the last two columns of Table S2. For the change of the stacking distance we find that it decreases by 0.26 (0.27) Å for the experimental and the relaxed unit cells of COF-1. Similarly to this small difference between the two unit cells also the total energies change by almost the same value of -0.866 (0.861) eV. Also for the interaction and the van der Waals energies such negligible differences between unit cells (i) and (ii) are found. For the electrostatic energy contribution, these differences become somewhat larger, but still remain small. To summarize, there are small numerical differences when looking at the trends of the energies. Nevertheless, the trends of the energies prevail.

*Table S2.* *Energies and structural parameters (displacement vectors, stacking distances) of COF-1 with the experimental and the optimized in-plane lattice parameters. z … stacking distance, $E_{total}$ … total energy per unit cell, $\Delta E_{int}$ … interaction energy, $\Delta E_{vdW}$ … vdW energy contribution, $\Delta E_{elstat}$ … electrostatic energy contribution, $\Delta E_{Pauli,orb}$ … Pauli repulsion with orbital rehybridization ; The stacking distances as well as the energies are given for the cofacial and the optimal arrangement for each COF.*

|  |  | COF-1 | COF-1 | Δ (optimal-cofacial) | |
| --- | --- | --- | --- | --- | --- |
| unit cell |  | experimental | relaxed | experimental | relaxed |
| z / Å | cofacial | 3.64 | 3.62 | -0.27 | -0.26 |
|  | optimal | 3.37 | 3.36 | | |
| $E_{total}$ / eV | cofacial | -70443.278 | -70443.846 | -0.866 | -0.861 |
|  | optimal | -70444.144 | -70444.707 | | |
| $\Delta E_{int}$ / meV | cofacial | -1954 | -2036 | -816 | -809 |
|  | optimal | -2770 | -2845 | | |
| $\Delta E_{vdW}$ / meV | cofacial | -2770 | -2863 | -532 | -527 |
|  | optimal | -3302 | -3390 | | |
| $\Delta E_{elstat}$ / meV | cofacial | -386 | -416 | -754 | -738 |
|  | optimal | -1140 | -1154 | | |
| $\Delta E_{Pauli,orb}$ / meV | cofacial | 1202 | 1243 | 470 | 456 |
|  | optimal | 1672 | 1699 | | |

### 2.3. Gaussian Process Regression: Model uncertainty and training data

As described in the main manuscript, the potential energy surface, PES, of COF-1 was obtained employing Gaussian Process Regression, GPR. A distinct advantage of GPR is that a model uncertainty, σ, can be obtained. As already described in the main manuscript our model (constant kernel times a radial basis function kernel (RBF) kernel) was trained on 80 randomly chosen data points and later 39 additional data points were added to the training data, in order to decrease the model uncertainty. These additional training data points were placed at positions with large model uncertainties. In Figure S2a the calculated potential energy surface (PES) can be seen, where all training data points (80 initial points plus 39 additional ones) are shown as black dots. The PES is the same as in Figure 2, albeit shown for a somewhat different plotting range. The main reason is to include all training data in the visualization. This is important for points with a radius/amplitude larger than 6 Å, as they were included in the training of our model. Figure S2b displays the corresponding model uncertainty. It can be seen that for the entire range of shifts that was considered, the model error is well below 60 meV. Especially for values of x and y smaller than 5 Å, this uncertainty is even well below 25 meV. Importantly, for displacements, where we find the global minimum of the total energy, the model uncertainty is even below 8 meV.

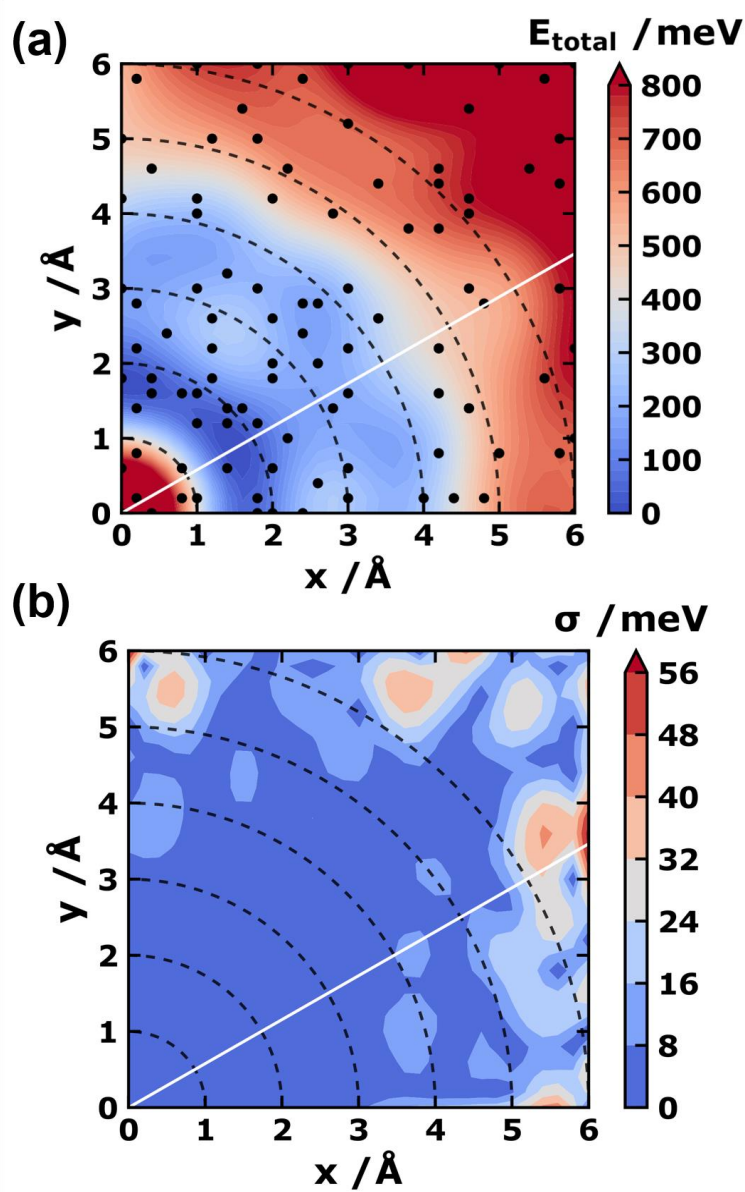

*Figure S2. Panel (a) shows the potential energy surface for COF-1 calculated using Gaussian Progress Regression together with the considered training data (black dots). In panel (b) one can see the model uncertainty within the considered range of xy-shifts. The black dash dotted*

*lines denote the amplitude of the displacements in terms of a radius. The white lines are located at 60° and indicate the six-fold symmetry of the system.*

## 2.4. Results for COF-1 for shifts perpendicular to a pore wall

In addition to the calculations of the interaction energy for interlayer shifts parallel to a pore wall, we also considered shifts of COF-1 layers along a direction perpendicular to the pore wall. For such shifts the interaction energy and the van der Waals energy have been determined for systems with a constant interlayer stacking distance and with optimized stacking distances. Considering the data for constant interlayer stacking distances in Figure S3a we find that the overall trend of the interaction energy for shifts parallel to the pore wall is also recovered here. Again the interaction energy $\Delta E_{int}$ has its highest value for the cofacial arrangement and exhibits a pronounced minimum at displacements around 1.75 Å. Comparing the total energies for the minima found for shifts parallel to the pore wall from the main manuscript and the direction considered here, it can be seen that the minimum for shifts perpendicular to the pore walls is ~40 meV higher in energy.

Furthermore, comparing the results for optimized interlayer stacking distances, we find that the overall trends are also similar. Again, the minimum found along direction **2** is slightly higher in energy (~50 meV) than the one found along direction **1**. Like for the constant interlayer stacking distance, a local minimum in the interaction energy is observed for displacements around 3.5 Å. However, this minimum is significantly higher in energy (~200 meV) than the observed minima at shifts of 1.75 Å.

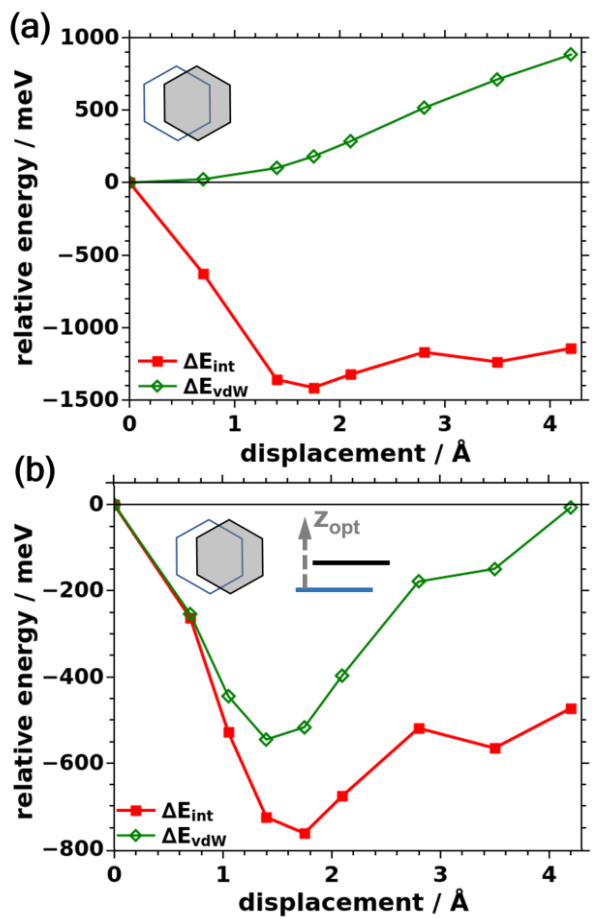

*Figure S3. Interaction and van der Waals energy for shifts of consecutive COF-1 layers along a direction perpendicular to the pore walls. In panel (a) the data for a system with constant interlayer stacking distance (3.328 Å) are shown. Panel (b) shows the energies for systems with stacking distances optimized at each displacement. The energies are given per unit cell containing two layers and are aligned to their respective values at cofacial arrangement. Energy values at 0.0 Å displacement for constant interlayer stacking distance: $\Delta E_{int}$=-1290 meV, $\Delta E_{int,elec}$=2257 meV, $\Delta E_{vdW}$=-3547 meV, $\Delta E_{electrostatic}$=-1344 meV, $\Delta E_{Pauli,orb}$=3601 meV, $E_{total}$=-70442.671 eV; Energy values at 0.0 Å displacement for optimized interlayer stacking distance: $\Delta E_{int}$=-1957 meV, $\Delta E_{int,elec}$=812 meV, $\Delta E_{vdW}$=-2769 meV, $\Delta E_{elstat}$=-385 meV, $\Delta E_{Pauli,orb}$=1197 meV, $E_{total}$=-70443.275 eV;*

## 2.5. Comparison of van der Waals corrections many body dispersion interactions vs. the Tkatchenko-Scheffler scheme

In this section we test, whether the observed trends (presented in the main manuscript) of the interaction energy as well as those of the individual energy contributions show any qualitative changes when treating the dispersion interactions by employing a many body dispersion (MBD) interaction scheme.[4–6] Within MBD, the system is described by a number of harmonic oscillators which are centered at the positions of the atoms. These oscillators are determined by polarizabilities which stem from the ground state electron density of the considered system. Based on these polarizabilities the MBD Hamiltonian is constructed. In order to determine the MBD energy correction, the obtained Hamiltonian is diagonalized. Such an MBD energy correction scheme is implemented in FHI-aims and has been employed to obtain data for a cofacial arrangement of COF-1 and for a shifted layer arrangement with a displacement of 1.75 Å. The in-plane lattice constants reported in literature have been used, see Methods section of main text. The interlayer stacking distance was optimized employing the MBD scheme and additionally we also kept it constant at the literature stacking distance. The obtained data can be found in Table S3 for the calculations with many body dispersion interactions and in Table S4 for the Tkatchenko-Scheffler scheme, vdW$^{TS}$.[7] What is important for the investigations presented in the main manuscript are the changes in the energies that are induced by interlayer shifts. Therefore, we can focus our considerations on these changes presented in Tables S3 and S4 (see the rows named "diff"). For the constant interlayer stacking distance we find that the deviations in the energy-shift differences between MBD and TS are negligible. For layer arrangements with optimized interlayer stacking distances, we find that MBD yields larger stacking distances compared to TS; also the difference between the stacking distance for cofacial and displaced structures are larger for MBD. Nevertheless, considering the changes in the interaction energy and the individual contributions, one finds that the overall trends are qualitatively the same for MBD and TS also when the stacking distance is optimized. This suggests that the trends presented in the main manuscript are not massively affected by the employed dispersion correction scheme.

*Table S3.* Interaction energy ad individual contributions when employing many body dispersion, MBD, corrections to the energy.

| constant interlayer stacking distance | | | | | |
|---|---|---|---|---|---|
| shift / Å | $\Delta E_{int}$ / meV | $\Delta E_{vdW}$ / meV | $\Delta E_{elstat}$ / meV | $\Delta E_{Pauli,orb}$ / meV | z / Å |
| 0.0 | -463 | -2720 | -1344 | 3601 | 3.328 |
| 1.75 | -1927 | -2565 | -1297 | 1935 | |
| diff / meV | -1464 | 155 | 47 | -1666 | 0.0 |
| optimized interlayer stacking distance | | | | | |
| 0.0 | -1393 | -1886 | -182 | 675 | 3.778 |
| 1.75 | -1973 | -2398 | -982 | 1407 | 3.415 |
| diff / meV | -580 | -512 | -800 | 732 | -0.363 |

*Table S4.* Interaction energy and individual contributions when employing the pairwise dispersion correction scheme by Tkatchenko and Scheffler, vdW$^{TS}$.

| constant interlayer stacking distance | | | | | |
|---|---|---|---|---|---|
| shift / Å | $\Delta E_{int}$ / meV | $\Delta E_{vdW}$ / meV | $\Delta E_{elstat}$ / meV | $\Delta E_{Pauli,orb}$ / meV | z / Å |
| 0.0 | -1290 | -3547 | -1344 | 3601 | 3.328 |
| 1.75 | -2751 | -3389 | -1297 | 1936 | |
| diff / meV | -1461 | 158 | 47 | -1665 | 0.0 |
| optimized interlayer stacking distance | | | | | |
| 0.0 | -1957 | -2769 | -385 | 1197 | 3.636 |
| 1.75 | -2760 | -3339 | -1207 | 1786 | 3.350 |
| diff / meV | -803 | -570 | -822 | 589 | -0.286 |

## 2.6. Electronic band structure evaluated with a hybrid functional

For the constant interlayer stacking distance we also calculated the electronic bands employing the hybrid functional HSE06.[8,9] There, we find that the evolutions of the valence band shows good agreement comparing PBE and HSE06 results (compare Figure 5 and Figure S4). Nevertheless, the bandwidths obtained with HSE06 are somewhat larger (VBW=2045 meV (1835 meV) for cofacial and VBW=1165 meV (1038 meV) for the shifted arrangement when calculated with the HSE06 (PBE) functional) and the effective mass is somewhat smaller for HSE06 ($m^*$=0.55 (0.85) $m_e$ for cofacial and $m^*$=1.13 (1.31) $m_e$ for the shifted arrangement again for HSE06 (PBE)).

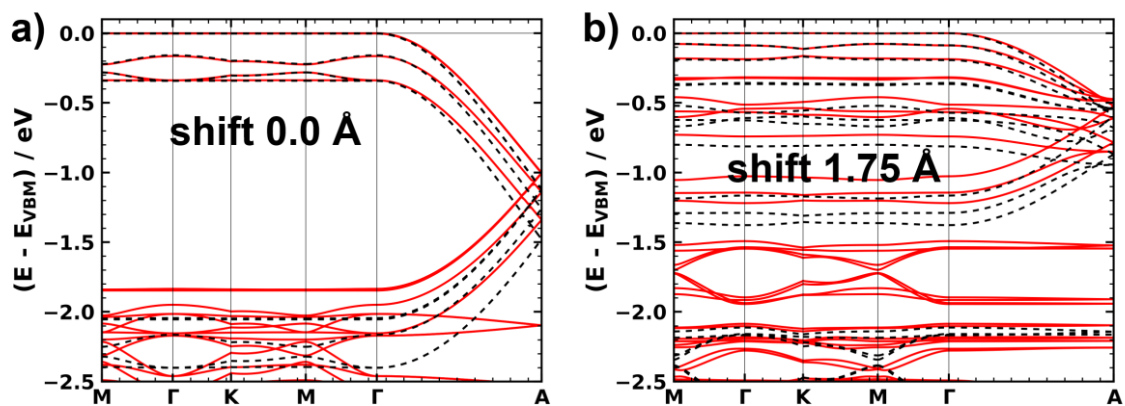

*Figure S4. Electronic band structure of COF-1 for cofacial (a) and minimum arrangements (b) shifted by 1.75 Å parallel to a pore wall for constant interlayer stacking distance. The band structures have been calculated using the PBE (solid red lines) functional and the hybrid HSE06 (dashed black lines) functional.*

## 2.7. Distance weighted histograms for COF-1

For rationalizing the evolution of the vdW interactions reported in the main manuscript we calculated histograms that show the number of atoms within a certain distance interval, i.e. that are within a sphere of a particular radius. This number of atoms was then weighted with the individual distance to the power of -6. This $1/R^6$ weight is inspired by terms that are included in typical pairwise vdW corrections. In Figure S5 the data for COF-1 shifted along the edge of the pores and relaxed interlayer stacking distances is shown. Figure S6 shows the data for the system without relaxed stacking distance.

Figure S5 shows that when optimizing the stacking distance the low distance fraction increases for displacements up to 1.75 Å. This trend is perfectly in line with the vdW interactions becoming more attractive in this range of displacements. For constant stacking distance, one observes that for layer displacements up to 1.75 Å only minor changes in the histogram appear, which correlates with the almost constant vdW interaction for that range. For larger displacements significant changes are observed, which is again consistent with the vdW interactions showing larges changes for such layer arrangements.

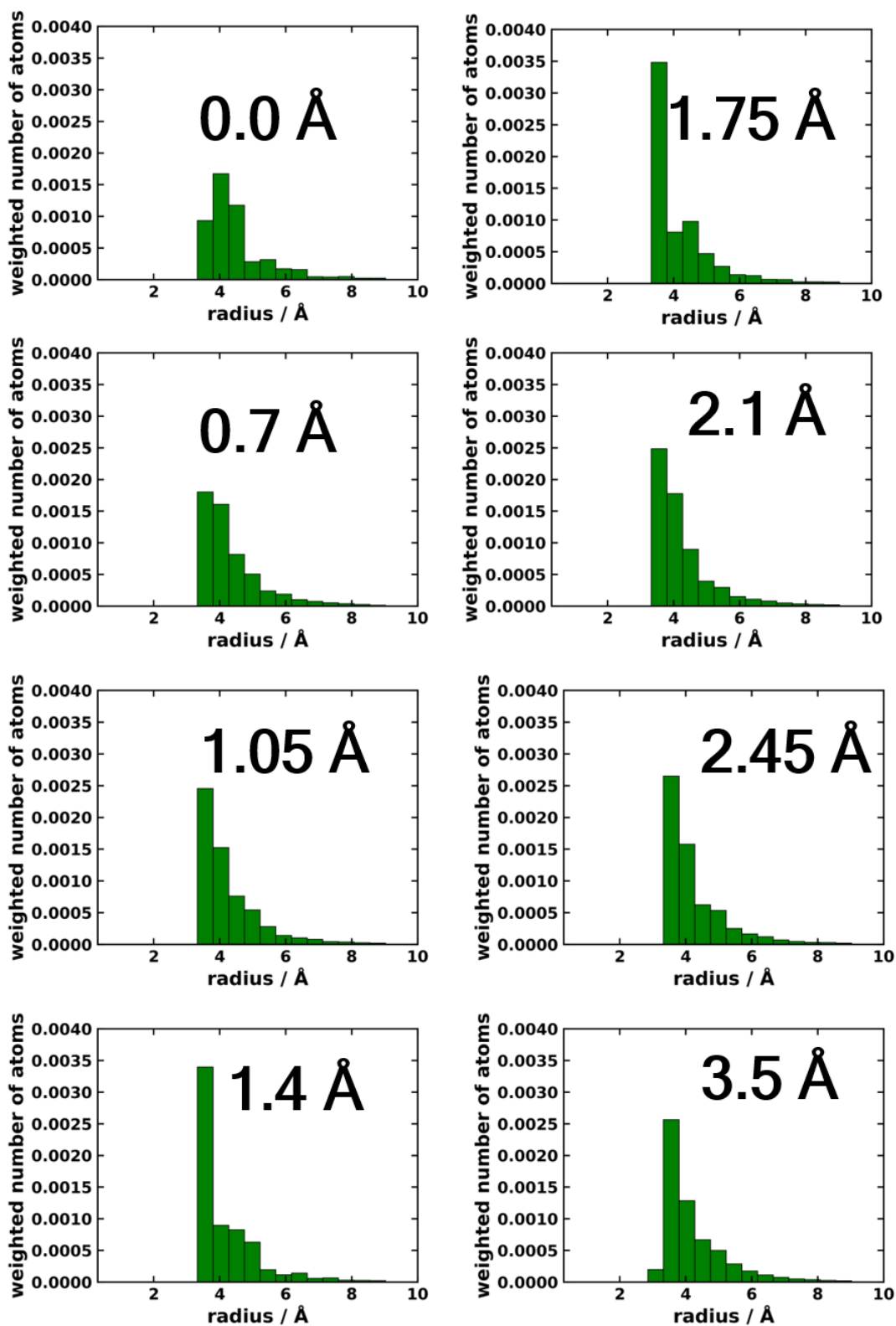

*Figure S5. Histogram showing the number of atoms found within a certain interlayer distance interval and weighted by this distance to the power of -6 (#atoms/($r^6$)) for COF-1 with optimized stacking distance. At each layer displacement such a histogram is created. One can see that the low distance contributions to this weighted number of atoms increase for displacements up to 1.75 Å, which is perfectly in line with the vdW interactions becoming more attractive in this region.*

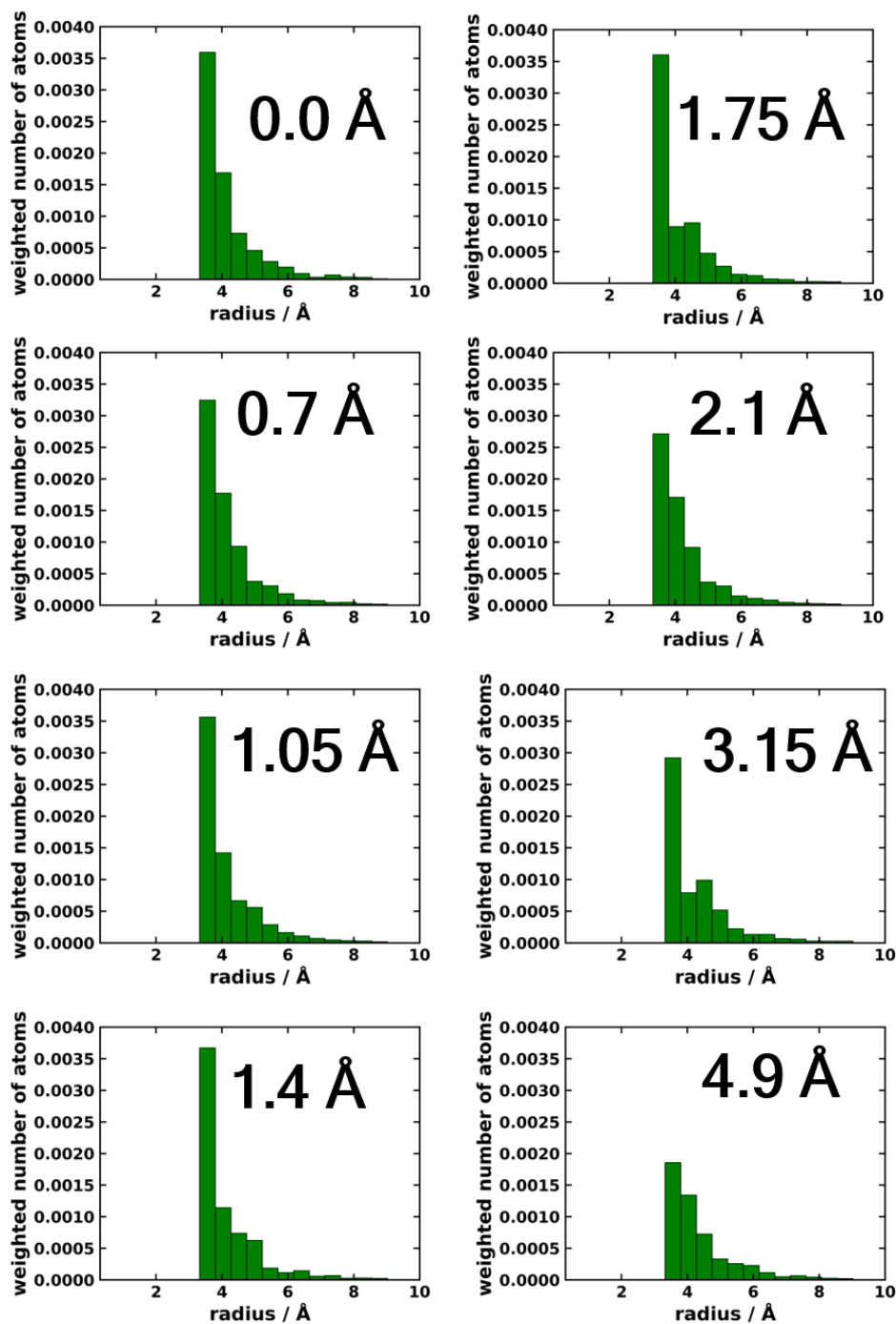

*Figure S6. Histogram showing the number of atoms found within a certain interlayer distance interval and weighted by this distance to the power of -6 (#atoms/(r^6)) for COF-1 with constant stacking distance. At each layer displacement such a histogram is created. For layer displacements up to 1.75 Å one can sees only little changes in the histogram, which correlated with the almost constant vdW interaction for that range. For larger displacements, significant changes are observed, again perfectly consistent with the vdW showing large changes for such layer arrangements.*

## 2.8. Comparison of total energy and interaction energy for COF-1

In Figure S7 one can see the evolution of the total energy and that of the interaction energy for COF-1 shifted along direction **1** with optimized stacking distances for each displacement. Both energies are aligned to their respective values for the cofacial arrangement. One can see that these energies essentially evolve in parallel and that only minor numerical differences occur. The reason why these energies do not coincide is that for each displacement the stacking distance and, thus, the unit cell vector along that direction, changes and so also the energies of fragments A and B (entering the determination of $\Delta E_{int}$) vary. Nevertheless, two energy curves show a excellent qualitative agreement.

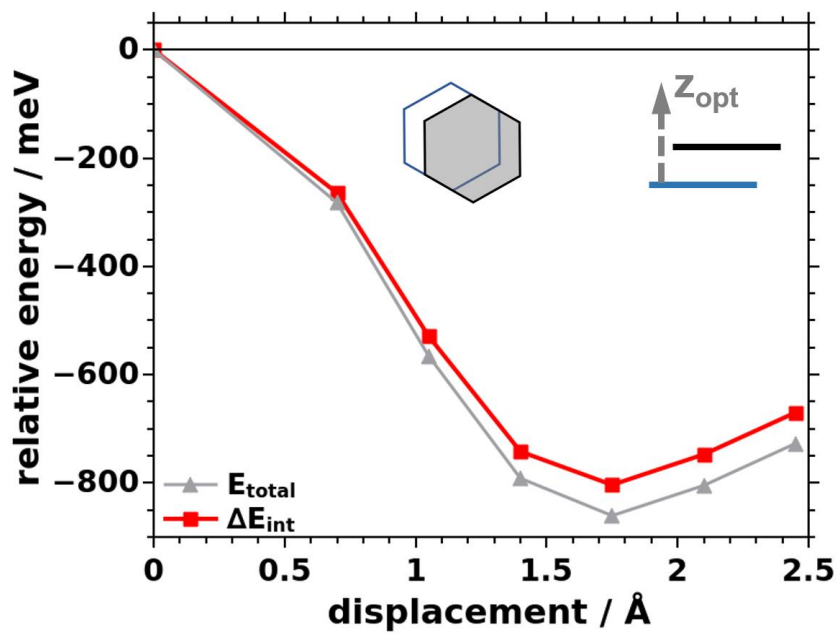

*Figure S7. Comparison of the evolution of the total energy (grey triangles and line) and the interaction energy (red squares and line) for COF-1 shifted along direction **1** with optimized stacking distances. Energies at cofacial arrangement: $\Delta E_{int}$=-1957 meV, $E_{total}$=-70443.275 eV;*

## 2.10. Relative energies of COF-5

In Figure S9 one can see the relative evolution of the energies of COF-5 as a function of displacements along a shift direction parallel to the edges of the pore, analogous to direction **1** of COF-1. For each displacement the stacking distance of consecutive COF-5 layers was optimized. Considering the evolution of the interaction energy we find that it exhibits a minimum at a displacement of around 1.5 Å, which is similar to COF-1. For both COFs, cofacial arrangements are energetically unfavorable and driving forces exist pushing these systems towards shifted layer arrangements. Decomposing the interaction energy into individual contributions comprising vdW interactions, electrostatic interactions and Pauli repulsion plus orbital rehybridization we find that their evolution shown in Figure S9b shows again very similar behavior compared to COF-1 (see Figure 6 of the main manuscript). Electrostatic and vdW interactions become more attractive upon layer displacements up to 1.5 Å and then, for larger displacements they become weaker again. The repulsion term (Pauli repulsion plus orbital rehybridization), on the contrary, gets more repulsive in the range of displacements where vdW and electrostatic contributions became more attractive. The sum of the changes in the vdW and the electrostatic interactions are larger than those of the repulsion, thus, they determine the formation of the minimum at the shifted layer arrangement of 1.5 Å.

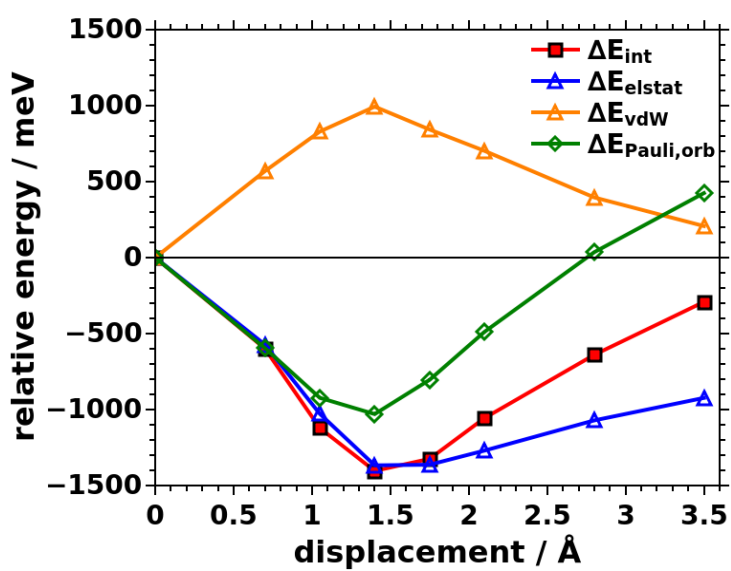

*Figure S9. (a) Relative energies of COF-5 as a function of the displacement for shifts along a direction parallel to one of the pore walls. The interlayer stacking distance was optimized at each displacement. Absolute energy values at 0.0 Å displacement: $\Delta E_{int,elec}$=1779 meV, $\Delta E_{vdW}$=-8483 meV, $\Delta E_{elstat}$=-1446 meV, $\Delta E_{Pauli,orb}$=3225 meV, $\Delta E_{int}$=-5238 meV*